\documentclass{article} 

\usepackage{lineno,hyperref}
\modulolinenumbers[5]

\usepackage{cite}

\usepackage{appendix}
\usepackage{amsmath,amssymb}
\usepackage{graphicx}
\usepackage{dcolumn}
\usepackage{booktabs}
\usepackage{float}
\usepackage{multirow}
\usepackage{bm}
\usepackage{hyperref}
\usepackage{braket}
\usepackage{soul}
\usepackage{stackengine}

\numberwithin{equation}{section}
\numberwithin{figure}{section}

\usepackage{xcolor}     

\definecolor{orcidlogocol}{HTML}{A6CE39}%

\hypersetup{
    colorlinks=true,
    linkcolor=blue,
    filecolor=blue,    
    citecolor=blue,
    urlcolor=blue,
    pdfpagemode=FullScreen,
    }

\begin{document}
\begin{center}
{\Large \bf 
Scalaroca stars: coupled scalar--Proca solitons
}
\vspace{0.8cm}
\\
{Alexandre M. Pombo$^1$,
João M. S. Oliveira$^2$, 
and  
Nuno M. Santos$^{3,4}$
\\
\vspace{0.3cm}
$^1${\small CEICO, Institute of Physics of the Czech Academy of Sciences, Na Slovance 2, 182 21 Praha 8, Czechia}\\
\vspace{0.3cm}
$^2${\small Centro de Matemática, Universidade do Minho, 4710-057 Braga, Portugal}\\
\vspace{0.3cm}
$^3${\small Departamento de F\'\i sica,
Instituto Superior T\'ecnico - IST, Universidade de Lisboa - UL,} 
\\ {\small Avenida
Rovisco Pais 1, 1049-001 Lisboa, Portugal}\\
\vspace{0.3cm}
$^4${\small Departamento de Matem\'atica da Universidade de Aveiro and } \\ {\small  Centre for Research and Development  in Mathematics and Applications (CIDMA),} \\ {\small    Campus de Santiago, 3810-183 Aveiro, Portugal} \\
\vspace{0.3cm}
}
\end{center}
\begin{abstract}
We construct and explore the physical properties of \textit{scalaroca stars}: spherically symmetric solitonic solutions made of a complex scalar field $\Phi$ and a complex Proca field $A^\mu$. We restrict our attention to configurations in which both fields are in the fundamental state and possess an equal mass, focusing on the cases when (\textit{i}) the scalar and Proca fields are (non--linearly) super--imposed and do not interact with each other; and (\textit{ii}) the scalar and Proca fields interact through the term $\alpha |\Phi| ^2 A^\mu A_\mu$. The solutions are found numerically for the non--interacting case ($\alpha=0$) as well as for both signs of the interaction coupling constant $\alpha$. While pure (\textit{\textit{i.e.}} single--field) Proca/scalar boson stars are the most/least massive for weakly--interacting fields, one can obtain more massive solutions for a sufficiently strong interaction. Besides, in the latter case, solutions can be either in a synchronized state -- in which both fields have the same frequency -- or in a non--synchronized state. In addition, we observe that the coupling between the two fields allows solitonic solutions with a real scalar field. We further comment on the possibility of spontaneous scalarization and vectorization of the interacting solitonic solution.
\end{abstract}

\tableofcontents

\section{Introduction}\label{sec:1}
%
	With the rise of gravitational--wave detections (led by the LIGO--Virgo collaboration~\cite{LIGOScientific:2016vbw,10.1063/1.882861,LIGOScientific:2014pky}) and the shadow imaging of supermassive objects (led by the Event Horizon Telescope collaboration~\cite{EventHorizonTelescope:2019dse,EventHorizonTelescope:2022wkp}), together with the various unsolved problems in fundamental physics, like the nature of dark matter and dark energy, the search for exotic objects is at full throttle. There is a plethora of hypothetical exotic compact objects, ranging from horizonless configurations to alternative black holes. If there happens to be enough evidence for any of these exotic objects or modifications to general relativity, it strongly suggests the existence of a new fundamental particle and its corresponding field. Some of these hypothetical particles are well--motivated candidates for physics beyond the Standard Model \cite{Freitas:2021cfi}.

	Over the past few decades several hypothetical models have been put forward and explored. The most common models feature a scalar field which, under specific conditions, give rise to boson stars \cite{Kaup:1968zz,Ruffini:1969qy,1968PhRv..168.1445F,Liebling:2012fv} and hairy black holes \cite{Gibbons:1987ps,Herdeiro:2015waa}. Boson stars (BSs)~\cite{Jetzer:1991jr} are of particular interest to this work (see \cite{Schunck:2003kk,Herdeiro:2017fhv,Herdeiro:2019mbz} for a comprehensive review).

	In a nutshell, scalar BSs (SBSs) are everywhere regular, localized, self--gravitating solutions of the minimally--coupled Einstein--Klein-Gordon system. In the simplest case, SBSs are lumps of a massive complex scalar field. Self--interacting configurations can also be considered~\cite{Colpi:1986ye,Lynn:1988rb,Schunck:1996he,Yoshida:1997qf,Astefanesei:2003qy,Schunck:2003kk,Liebling:2012fv,Grandclement:2014msa,Alcubierre:2018ahf,Guerra:2019srj,Delgado:2020udb,Herdeiro:2020kvf,Herdeiro:2021lwl}.

	Unlike ordinary stars, in the simplest models, BSs do not interact with the Maxwell field, thus being transparent and invisible. Still, when immersed in an environment with ordinary matter, they can be compact enough to bend light due to the gravitational pull~\cite{Cunha:2015yba}, creating an empty region resembling a shadow of a BH event horizon. However, the accreted matter would be visible in their interior~\cite{Herdeiro:2021lwl}. 

	While the more straightforward mathematical structure of scalar fields is appealing, it does not exhaust all possibilities. A commonly used alternative is the massive vector (or Proca) field, and their version of self--gravitating stars: Proca stars (PSs)\footnote{Rotating black holes in equilibrium with a vector field are also possible\cite{Herdeiro:2016tmi,Santos:2020pmh,Heisenberg:2017xda}.} \cite{Brito:2015pxa} (see also~\cite{Herdeiro:2017fhv,Herdeiro:2019mbz,Minamitsuji:2018kof,Herdeiro:2020jzx}). Proca fields can also be subject to a self--interacting potential, much like their scalar counterparts~\cite{Minamitsuji:2018kof}. However, unlike scalar fields, self--interacting Proca fields are prone to ghost instabilities and their time evolution can even break down due to the loss of hyperbolicty \cite{Clough:2022ygm,Mou:2022hqb,Coates:2022qia,Coates:2023dmz}. 

	In general, BSs have been shown to possess a stable branch~\cite{Cunha:2017wao} dominated by the mass term and common to every BS configuration. However, only diluted configurations exist there, making it hard to describe astrophysical observations with them -- see~\cite{Herdeiro:2021lwl,Pombo:2023ody}. More compact BSs can be found in theories featuring \emph{multiple} bosonic fields.

	While single--field BSs have been thoroughly studied in the literature (some multi--state configurations have been considered in \cite{Bernal:2009zy}), multi--field configurations are still poorly explored. Examples of multi--field BSs include $\ell$--boson stars \cite{Alcubierre:2018ahf} and Higgs--Proca stars~\cite{Dzhunushaliev:2021vwn,Dzhunushaliev:2019ham,Herdeiro:2023lze}.\footnote{A work in the same spirit but with a scalar and a Dirac field has recently been published \cite{Liang:2022mjo}, however, without field interactions.}$^,$\footnote{Right before our work appeared, other authors have reported the existence of hybrid scalar--Proca stars in a similar but simpler model~\cite{Ma:2023vfa}.}

	We study a model with a massive complex scalar field and a massive complex vector field, both minimally--coupled to Einstein's gravity but non--minimally coupled to each other in such a way that their effective mass can change. We study the two fields' resulting superposition (with and without the non--minimal interaction) for specific ratios of their frequencies while keeping the mass of the individual fields particle unaltered. We do not consider higher-order self-interaction terms for the fields, so they are not subject to the pathologies mentioned above.

	We study these self--gravitating scalar--Proca stars (SPSs, or \textit{scalaroca stars}) and compare them to single--field BSs, namely SBSs and PSs.\footnote{While the term ``boson star" usually refers to scalar fields only, here it will denote self--gravitating bosonic solitons in general. In other words, in this work, SBSs, PSs and SPSs are all BSs.} In particular, we consider ($i$) non--interacting (or purely gravitational) configurations, in which the interaction between the fields is exclusively ruled by gravity; ($ii$) interacting configurations, whose equilibrium follows from a balance between gravity and the direct interaction between the fields.

	In \autoref{sec:2} we introduce the theoretical framework, namely the action and equations of motion (\autoref{sec:2.1}), the ans\"{a}tze (\autoref{sec:2.2}) and the boundary conditions (\autoref{sec:2.3}), and explore some features of the theory (\autoref{sec:2.4}). In \autoref{sec:3} we present and discuss the results for the four aforementioned cases. The conclusion and remaks on future work can be found in \autoref{sec:4} .

	Throughout the paper, $4\pi G=1=4\pi\epsilon_0$. The signature of the spacetime is $(-,+,+,+)$. In this work one is solely interested in spherical symmetry and the metric matter functions are only radially dependent. For notation simplicity, after being first introduced, the functions' radial dependence is omitted, \textit{e.g.} $X(r)\equiv X$, and  $X' \equiv dX/dr$. At last, an overbar denotes complex conjugation, \textit{e.g.} $\bar{X}$.
%
\section{Framework}\label{sec:2}
%
    \subsection{Action and equations of motion}\label{sec:2.1}
%
	Consider the field theory of a complex scalar field, $\Phi$, non--minimally coupled to a complex vector field, $A^\alpha$, defined by the action
		\begin{equation}
		 \mathcal{S} = \int \text{d}^4 x \sqrt{-g} \Bigg[\frac{R}{4}-\Big(\Phi_{,\mu}\bar{\Phi}^{,\mu}+\mu_\Phi^2|\Phi|^2\Big) -\Big(\frac{1}{4}F_{\mu\nu}\bar{F}^{\mu\nu}+\frac{1}{2}\mu_A^2\textbf{A}^2\Big)-\alpha |\Phi|^2\textbf{A}^2\Bigg]\ ,
		 \label{eq:2.1}
		\end{equation}
where $g_{\mu\nu}$ is the spacetime metric, with determinant $g$ and Ricci scalar $R$, $F=\text{d}A$ is the Maxwell tensor, $\mathbf{A}^2\equiv A^\mu\bar{A}_\mu$, $\mu_\Phi$ and $\mu_A$ are the scalar and vector field bare mass, respectively, and $\alpha$ is the interaction coupling constant.

	The resulting equations of motion are
	\begin{subequations}
	\begin{align}
		\label{eq:2.2a}
		&R_{\mu\nu}-\frac{1}{2}g_{\mu\nu}R =  2\, T_{\mu\nu}\ ,\\
		\label{eq:2.2b}
		&\nabla_\nu F^{\mu\nu}+\hat{\mu}_A^2A^\mu=  0\ ,\\
		\label{eq:2.2c}
		&\left(\Box-\hat{\mu}_\Phi^2\right)\Phi=0\ ,
	\end{align}
	\end{subequations}
where $\hat{\mu}_\Phi^2\equiv\mu_\Phi^2+\alpha\textbf{A}^2/2$ and $\hat{\mu}_A^2\equiv\mu_\Phi^2+\alpha|\Phi|^2$ are the effective masses of the scalar and vector fields, respectively. They coincide with their bare masses when $\alpha=0$. Note that the four--divergence of~\eqref{eq:2.2b} implies the \textit{modified} Lorenz condition $\nabla_\mu\big(\hat{\mu}_A^2A^\mu\big)=0$.

	The stress--energy tensor $T_{\mu\nu}$ can be written as
	\begin{equation}
		 T_{\mu\nu}= T_{\mu\nu} ^{(\Phi)}+T_{\mu \nu} ^{(A)}+T_{\mu \nu} ^{(\alpha)}\ ,
		\end{equation}
where
	\begin{subequations}
	\begin{align}
		\label{eq:2.4a}
		 &T_{\mu\nu}^{(\Phi)}=(\Phi_{,\mu}\bar{\Phi}_{,\nu}+\Phi_{,\nu}\bar{\Phi}_{,\mu})-g_{\mu\nu}\left(\Phi_{,\lambda}\bar{\Phi}^{,\lambda}+\mu_\Phi^2|\Phi|^2\right)\ ,\\
		 \label{eq:2.4b}
		 &T_{\mu\nu}^{(A)}=\frac{1}{2}\left(F_{\mu\gamma}\bar{F}_{\nu\lambda}+F_{\nu\gamma}\bar{F}_{\mu\lambda}\right)g^{\lambda\gamma}+\frac{\mu_A^2}{2}(A_\mu\bar{A}_\nu+A_\nu\bar{A}_\mu)-g_{\mu\nu}\left(\frac{1}{4}F_{\lambda\gamma}\bar{F}^{\lambda\gamma}+\frac{\mu_A^2}{2}\textbf{A}^2\right)\ ,\\
		 \label{eq:2.4c}
		&T_{\mu \nu}^{(\alpha)}=\alpha \Big[\mu_A^2|\Phi| ^2 (A_\mu\bar{A}_\nu+A_\nu\bar{A}_\mu)-g_{\mu\nu}|\Phi|^2\textbf{A}^2\Big]\ ,
	\end{align}
	\end{subequations}
are the scalar, vector and scalar--vector interaction contributions, respectively. 

	The action in \eqref{eq:2.1} possesses two global $\textbf{U}(1)$ symmetries, being invariant under the transformations $\Phi\rightarrow e^{i\chi}\Phi$ and $A_\mu\rightarrow e^{i\xi}A_\mu$, where $\chi$ and $\xi$ are constant. This implies the existence of two conserved four--currents
	\begin{equation}
		\label{eq:2.5}
		 j_\Phi^\mu=-i(\bar{\Phi}\Phi^{,\mu}-\Phi\bar{\Phi}^{,\mu})\ ,\qquad \qquad j_A^\mu=\frac{i}{2}\left(\bar{F}^{\mu\nu}A_\nu-F^{\mu\nu}\bar{A}_\nu\right)\ ,
	\end{equation}
which are conserved, \textit{i.e.} $\nabla_\mu j_\Phi^\mu=0$ and $\nabla_\mu j_A^\mu=0$. The corresponding Noether charges are obtained by integrating the timelike component of the four--currents on a spacelike surface $\Sigma$,
	\begin{equation}
		\label{eq:2.6}
		Q_\Phi=\int_\Sigma\text{d}^3x~j_\Phi^0\ ,\qquad \qquad	Q_A=\int_\Sigma\text{d}^3x~j_A^0\ .
	\end{equation}
Upon quantization, $Q_\Phi$ and $Q_A$ are nothing but the number of scalar and vector particles, respectively. $Q\equiv Q_\Phi+Q_A$ is thus the total number of particles.
    The Komar mass reads
    \begin{align}
    		\label{eq:2.7}
        M=\int_\Sigma \text{d}V R_{\mu\nu}n^\mu\xi^\nu=2\int_\Sigma\text{d}V\left(T_{\mu\nu}-\frac{1}{2}g_{\mu\nu}{T^\lambda}_\lambda\right)n^\mu\xi^\nu\ ,
    \end{align}
where $\Sigma$ is an asymptotically--flat spacelike hypersurface, $n^\alpha$ is a future--pointing unit normal to $\Sigma$, and $\text{d}V$ is the $3$--volume form induced on $\Sigma$.
%
    \subsection{Ans\"{a}tze}\label{sec:2.2}
%
	Any static, spherically--symmetric solution to the equations of motion can be cast in the form
		\begin{align}
		  \text{d}s^2=-\sigma(r)^2N(r)\text{d}t^2+\frac{\text{d}r^2}{N(r)}+r^2\left(\text{d}\theta^2+\sin^2\theta\text{d}\varphi^2\right)\ ,
		  \quad
		  N(r)\equiv1-\frac{2m(r)}{r}\ ,
		  \label{E2.17}
		\end{align}
in Schwarzschild-like coordinates $(t,r,\theta,\varphi)$, where $m$ is the Misner--Sharp mass function~\cite{Misner:1964je}. Note that $\delta=\log(\sigma\sqrt{N})$ is the redshift function. For both matter fields, one considers an ansatz with an harmonic time-dependence that makes the stress energy tensor time-independent. The ansatz for the scalar field  is
		\begin{align}
		 \Phi(t,r)=e^{-i\omega t}\phi(r)\ ,
		\end{align}	 
where $\phi$ is the scalar field amplitude, which depends on the radial coordinante only, and $\omega$ is the field's frequency. For the vector field, the ansatz reads
		\begin{align}
		 A_\alpha(t,r)\text{d}x^\alpha=e^{-i \gamma t}\big[f(r)\text{d}t+ig(r)\text{d}r\big] \ ,
		\end{align}
where $f$ and $g$ depend on the radial coordinate only and $\gamma$ is the associated frequency. Without loss of generality, both $\omega$ and $\gamma$ are taken to be positive. 

	The equations of motion restricted to the ans\"{a}tze
	\begin{subequations}
		\begin{align}
    		 &\frac{1}{r^2\sigma}(r\sigma N)'-\frac{1}{r^2}+2\mu_\Phi^2\phi^2+\frac{\sigma^2N^2g^2}{\gamma^2}\big(\mu_A^2+\alpha\phi^2\big) ^2=0\ ,\\
    		 &\frac{\sigma'}{\sigma}-r\left[\left(\mu_A^2+\alpha\phi^2\right)\left(\frac{f^2}{\sigma^2N^2}+g^2\right)+\frac{2\omega^2\phi^2}{\sigma^2N^2}+2\phi'^2\right]=0\ ,\\
    		 &\frac{1}{r^2\sigma}(r^2\sigma N\phi')'+\left[\frac{\omega^2}{\sigma^2N}-\mu_\Phi^2-\frac{\alpha}{2}\left(Ng^2-\frac{f^2}{\sigma^2N}\right)\right]\phi=0\ ,\label{E2.20}\\
    		 &f'-\gamma g+\frac{\sigma^2N}{\gamma}\big( \mu_A^2+\alpha\phi^2\big) g=0\ ,
    		 \label{E2.21}
		\end{align}
		\label{eq:2.14}
	\end{subequations}
	the modified Lorenz condition reads\footnote{By multiplying \eqref{eq:2.12} by $r^2(\mu_A^2+\alpha\phi^2)$ and integrating from $r=0$ to infinity, one obtains
			\begin{align*}
			 \int_{0}^\infty\text{d}r\,\frac{r^2}{\sigma N}(\mu_A^2+\alpha\phi^2)f=0\ .
			\end{align*}
	Assuming $\alpha>0$, this equality is only satisfied if $f$ changes sign at least once (and thus have at least one node).}.
		\begin{align}
		 \frac{1}{r^2(\mu_A^2+\alpha\phi^2)}\Big[r^2\sigma N\big(\mu_A^2+\alpha\phi^2\big)g\Big]'+\frac{\gamma}{\sigma N}f=0\ .
		 \label{eq:2.12}
		\end{align}
	Note that the ordinary differential equation (ODE) for $\phi$ is of second--order, whereas those for $N$, $\sigma$ and $f$ are of first--order\footnote{The matter function $g$ can be expressed in terms of $N$, $\sigma$, $\phi$, $f$ and its first derivative.}.

%
    \subsection{Boundary conditions and physical quantities}\label{sec:2.3}

	As for the inner boundary conditions, smoothness requires the functions $\{N,\sigma,\phi,f,g\}$ to have a regular Taylor series at $r=0$. In fact, it can be shown that
	\begin{subequations}
	\label{eq:2.13}
	\begin{align}
		 &N(r)=1-\frac{2}{3}\frac{\phi_0^2}{\sigma_0^2}\left[\omega^2+\mu_\Phi^2\sigma_0^2+\frac{f_0^2}{2\phi_0^2}\big(\mu_A^2+\alpha\phi_0^2\big)\right]r^2+\ldots\ ,\\
		 &\sigma(r)=\sigma_0+\frac{\phi_0^2}{\sigma_0}\left[\omega^2+\frac{f_0^2}{2\phi_0^2}\big(\mu_A^2+\alpha\phi_0^2\big)\right]r^2+\ldots\ ,\\
		 &\phi(r)=\phi_0-\frac{1}{6}\frac{\phi_0}{\sigma_0^2}\left(\omega^2-\mu_\Phi^2\sigma_0^2+\frac{\alpha}{2} f_0^2\right)r^2+\ldots\ ,\\
		 &f(r)=f_0\left[1+\frac{1}{6}\left(\mu_A^2+\alpha\phi_0^2-\frac{\gamma^2}{\sigma_0^2}\right)r^2+\ldots\right]\ ,\\
		 &g(r)=-\frac{\gamma}{3}\frac{f_0}{\sigma_0^2}r+\ldots\ ,		 
	\end{align}
	\end{subequations}
where $\sigma_0\equiv\sigma(0)$, $\phi_0\equiv\phi(0)$ and $f_0\equiv f(0)$. Without loss of generality, one can assume that $\phi_0>0$ thanks to the $\mathbb{Z}_2$--symmetry of the scalar field. Note that $N'(0)=\sigma'(0)=\phi'(0)=f'(0)=0$.

	As for the outer boundary conditions, asymptotic flatness requires
	\begin{subequations}
	\begin{align}
    	 	 &\lim_{r\rightarrow\infty}N(r)=\lim_{r\rightarrow\infty}\sigma(r)=1\ ,\\
    	 	 &\lim_{r\rightarrow\infty}\phi(r)=\lim_{r\rightarrow\infty}f(r)=\lim_{r\rightarrow\infty}g(r)=0\ .
	\end{align}
	\end{subequations}
	More precisely, the asymptotic behaviour of the functions $\{N,\sigma,\phi,f,g\}$ is of the form
	\begin{subequations}
	\label{eq:2.15}
	\begin{align}
		\label{eq:2.13a}
		&N(r)=1-\frac{2M}{r}+\cdots\ ,\\
		&\sigma(r)=-\frac{c_0^2}{2}\frac{\mu_A^2\gamma^2}{(\mu_A^2-\gamma^2)^{3/2}}\frac{e^{-2r\sqrt{\mu_A^2-\gamma^2}}}{r}+\ldots\ ,\\	
		&\phi(r)=\phi_\infty\frac{e^{-r\sqrt{\mu_\Phi^2-\omega^2}}}{r}+\ldots\ ,\\
		&f(r)=f_\infty\frac{e^{-r\sqrt{\mu_A^2-\gamma^2}}}{r}+\ldots\ ,\\
		&g(r)=f_\infty\frac{\gamma}{\sqrt{\mu_A^2-\gamma^2}}\frac{e^{-r\sqrt{\mu_A^2-\gamma^2}}}{r}+\ldots\ ,
			\label{E2.25}
	\end{align}
	\end{subequations}
	where $c_0$, $\phi_\infty$ and $f_\infty$ are real constants.
	Restricted to the ans\"{a}tze, the conserved Noether charges in \eqref{eq:2.6} read
		\begin{align}
		 Q_\Phi=8\pi\int_0^{+\infty} \text{d}r~r^2\frac{\omega}{\sigma N}\phi^2\ ,
		 \quad
		 Q_A=-4\pi\int_0^{+\infty}\text{d}r~r^2\frac{g(f'-\gamma g)}{\sigma}\ ,
		\end{align}
and the Komar mass in \eqref{eq:2.7}  becomes
	\begin{align}
			 \label{E2.28}
		 M=4\pi\int_0^{\infty} dr~\frac{r^2}{\sigma N}\left[4\left(\omega^2-\frac{\mu_\Phi^2}{2}\sigma^2N\right)\phi^2+N\big( f'-\gamma g\big) ^2+2f^2\big( \mu_A^2+\alpha\phi^2\big) \right]\ .
	\end{align}
The mass $M$ can also be read off from the behaviour of the metric function $N$ at infinity, \eqref{eq:2.13a}. Note that the positive mass theorem is not violated if $\omega$ is set to $0$ in \eqref{E2.28}, as the coupling between the scalar and the Proca fields allows for a positive Komar mass regardless of the negative contribution of the scalar field.

	    At last, following \cite{Herdeiro:2021teo,Herdeiro:2022ids,Oliveira:2022hen}, applying Derrick's scaling argument, one can obtain the virial identity for the full model,
	\begin{align}
         \int_0 ^{+\infty} dr\ \frac{r^2}{N^2 \sigma}\Bigg[& 2\omega ^2 \phi ^2\big(1-4N\big)+6\, \mu_\Phi ^2 N^2 \sigma ^2\phi ^2-f^2\big(-1+4N\big)\big(\mu _A ^2+ \alpha \phi ^2\big)+\nonumber\\
        &g^2 N^2\Big[-3\gamma+\big(1+2N)\sigma^2\big(\mu _A ^2+ \alpha \phi ^2\big)\Big]
         +\big(4\gamma\, g -f'\big)N^2 f'+2N^2 \sigma ^2\phi'^2 \Bigg] = 0\ .
         \label{E2.36}
	\end{align} 
    Observe that while the computation was performed with the $m$ function, for notation simplicity we express the resulting identity with the $N$ function.	
	
%
    \subsection{Tachyonic instabilities and spontaneous bosonification}\label{sec:2.4}

	The model in \eqref{eq:2.1} is likely to feature (either scalar or vector) \textit{tachyonic instabilities} due to the presence of the interaction term $\alpha|\Phi|^2\mathbf{A}^2$. Once triggered, they may lead to the growth of the bosonic field, commonly known as \textit{spontaneous scalarization} (for spin--0 fields) and \textit{spontaneous vectorization} (for spin--1 fields). For simplicity, we will hereafter refer to this phenomenon as \textit{spontaneous bosonification}. 
	
	First proposed by Damour and Esposito-Farése~\cite{Damour:1992we,Damour:1993hw}\footnote{Reports of an earlier proposal by Zagaluer, back in 1992, also exist; however, due to some artificial considerations, it has been criticized ever since \cite{Zaglauer:1992bp}.}, scalarization occurs when, a non--trivial configuration of a scalar field with vanishing asymptotic behaviour is dynamically preferred. It is said to be spontaneous scalarized when such a scalar configuration occurs without an inducing external perturbation (hence the name). The spontaneous scalarization of SBS has already been reported in \cite{Brihaye:2019puo}\footnote{The same phenomena also occurs for neutron stars \cite{Damour:1993hw} and black holes \cite{Doneva:2017bvd,Cardoso:2013opa,Herdeiro:2019yjy,Silva:2017uqg,Herdeiro:2018wub,Herdeiro:2021vjo,LuisBlazquez-Salcedo:2020rqp,Fernandes:2019kmh,Fernandes:2019rez,Astefanesei:2019pfq}.}. As is the case in our model, the same can also occur in the presence of a vector field \cite{Ramazanoglu:2017xbl,Oliveira:2020dru,Kase:2020yhw} and even higher rank tensor fields \cite{Ramazanoglu:2019gbz}.
	
	It is clear from \eqref{eq:2.2b} that the scalar field is prone to such instabilities when $\hat{\mu}_\Phi^2$ becomes negative:
	\begin{subequations}
	\begin{align}
		\mu_\Phi^2 &< -\alpha |\textbf{A}^2|/2 \qquad\qquad \mathrm{for}\; \alpha >0 , \; \textbf{A}^2<0\ , \\
		\mu_\Phi^2 &< |\alpha| \textbf{A}^2/2  \qquad\qquad \;\;\;\mathrm{for}\; \alpha <0 , \; \textbf{A}^2>0\ .
	\end{align}
	\end{subequations}
	
	which correspond to when the effective mass $\hat{\mu}_\Phi$ of the scalar field is imaginary. All these branches of solution should be distinct. Note that for the $\mathbf{A}^2 <0$ case one requires $|g^{tt}||A_t|^2 > g^{rr} |A_r|^2$.
	
	As for the Proca field, tachyonic instabilities may arise when the eigenvalues of the effective mass matrix become negative. Such matrix can be obtained by rewriting the Proca equation \eqref{eq:2.2c} in the form $g_{\mu\nu} \nabla^\mu\nabla^\nu A_\sigma - \mathcal{M}_{\sigma\lambda} A^\lambda = 0$. Here,
	\begin{align}
		&\mathcal{M}_{\sigma\lambda} = R_{\sigma\lambda} + \bar{\mu}_{A}^2 g_{\sigma\lambda} -\nabla_\sigma\nabla_\lambda \ln(\hat{\mu}_{A}^2) \label{Meff} \ .
	\end{align}
	
	An indepth study of spontaneous bosonification in model~\eqref{eq:2.1} is out of the scope of this work.
%
\section{Results}\label{sec:3}
%
\bigskip
	Observe that the system~\eqref{eq:2.14} is invariant under the transformations
\begin{subequations}	
	\begin{align}
		&\{\sigma,f,\omega,\gamma\}\rightarrow\lambda_a\{\sigma,f,\omega,\gamma\}\label{eq:3.1a}\ ,\\
		&\{r,m\}\rightarrow\lambda_b\{r,m\} \ ,
		\quad
		\{\mu_\Phi,\mu_A,\omega,\gamma\}\rightarrow\lambda_b^{-1}\{\mu_\Phi,\mu_A,\omega,\gamma\}\ ,
		\quad
		\alpha\rightarrow\lambda_b^{-2}\alpha\ ,\label{eq:3.1b}
	\end{align}
\end{subequations}	
where $\lambda_a,\lambda_b\in\mathbb{R}^+$. Equation~\eqref{eq:3.1b} leaves $\omega/\mu_\Phi$ and $\gamma/\mu_A$ unchanged. It is thus convenient to set the mass of one of the fields equal to unity, $\mu_A=1$ (say), which yields $\lambda_b=1/\Tilde{\mu}_A$, and physical quantities will be scaled accordingly. While solutions with $\mu_A \neq \mu _\Phi$ are possible, here one sets $\mu\equiv\mu_A=\mu_\Phi=1$, and focuses on the effects of changing the coupling $\alpha$ \footnote{A change in $\mu_A/\mu_\Phi$ is expected to impact on the starting point and size of the domain of existence, which increases (decreases) as the ratio decreases (increases)~\cite{Liang:2022mjo}.}. 

	Solitonic solutions are characterized by three continuous parameters: the ADM mass, $M$, and the two oscillation frequencies of the matter fields, $\omega$ and $\gamma$, all expressed in units of $\mu$. For each value of $(\omega/\mu,\gamma/\mu,M\mu)$, there is a single solution in a certain three--dimensional domain. This family of solutions is only one amongst an infinite discrete set, labeled by two integers: the number of nodes in the radial direction $n_\phi$ and $n_f$ of the matter functions $\phi$ and $f$, respectively. Fundamental solutions, which minimize the total number of nodes $n_\phi+n_f$, are the main focus of the paper. Excited solutions are expected to exist, though.
	
	For a given $\{\omega,\gamma,\alpha\}$, the system is solved as a boundary value problem with the boundary conditions in \eqref{eq:2.13} and \eqref{eq:2.15}. The set of coupled ODEs is integrated numerically by using an in house developed parallelized, adaptative step--size Runge--Kutta method of order $5(6)$, with a local truncation error of $10^{-15}$. The pair $\{\phi_0,f_0\}$ which satisfies the boundary conditions at infinity is found by implementing a two--dimensional shooting method based on the secant algorithm, with a tolerance of $10^{-9}$. The physical accuracy of the solutions (required to be at least $10^{-5}$) is monitored by checking: ($i$) the virial identity in \eqref{E2.36}; ($ii$) the difference between the Komar mass in \eqref{E2.28} and the asymptotic value of $m$.

	The domain of existence of single--field BSs has a spiral shape in a mass--frequency diagram, as shown in ~\autoref{F3.2} (left). BSs exist in a finite range of frequencies and masses: while SBSs can oscillate slower, $\min(\omega)<\min(\gamma)$, PSs can be heavier, $\max(M_\text{SBS})<\max(M_\text{PS})$. SPSs lie along line segments connecting the two spirals, and their endpoints correspond to single--field (pure) BSs, \textit{i.e.} either a SBS ($f_0=0$) or a PS ($\phi_0=0$). As one moves along the line segment towards the PS (say), $f_0$ increases and $\phi_0$ decreases -- \autoref{F3.2} (right).

%
    \subsection{Minimal coupling ($\alpha=0$)}\label{sec:3.1}
\begin{figure}[H]
    		\centering
		\includegraphics[scale=0.7]{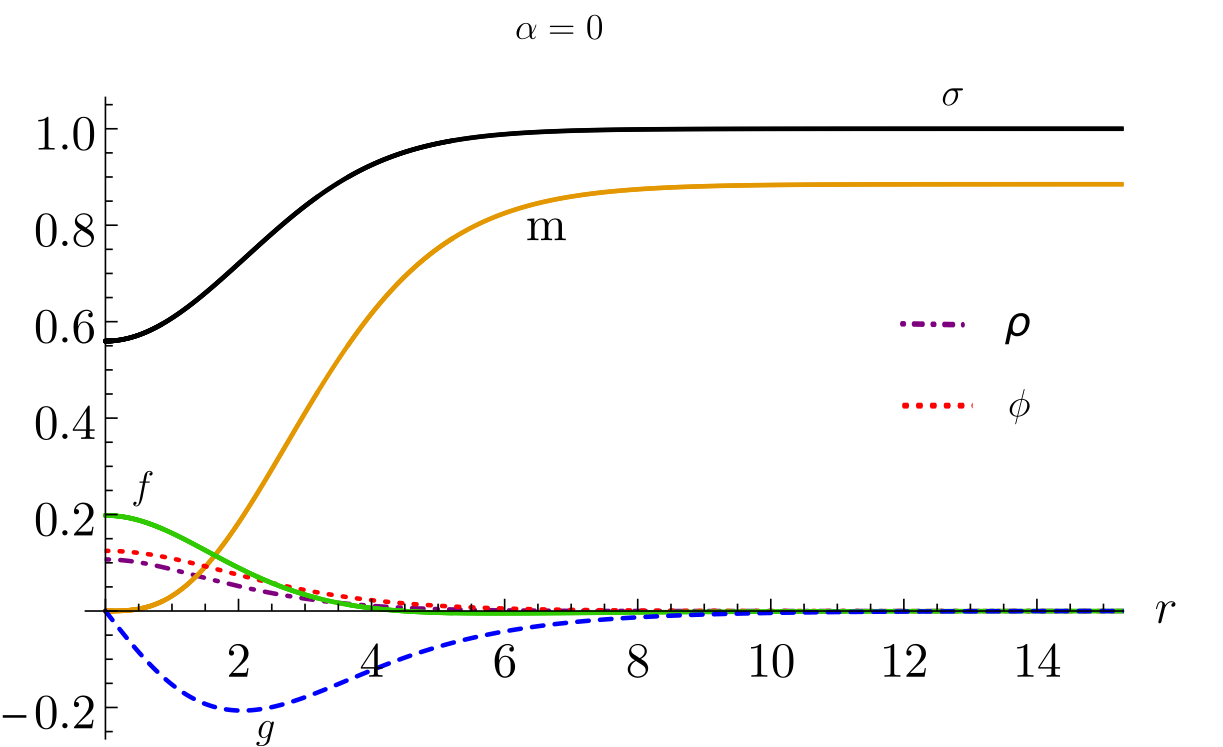}
		\hfill
   		\includegraphics[scale=0.4]{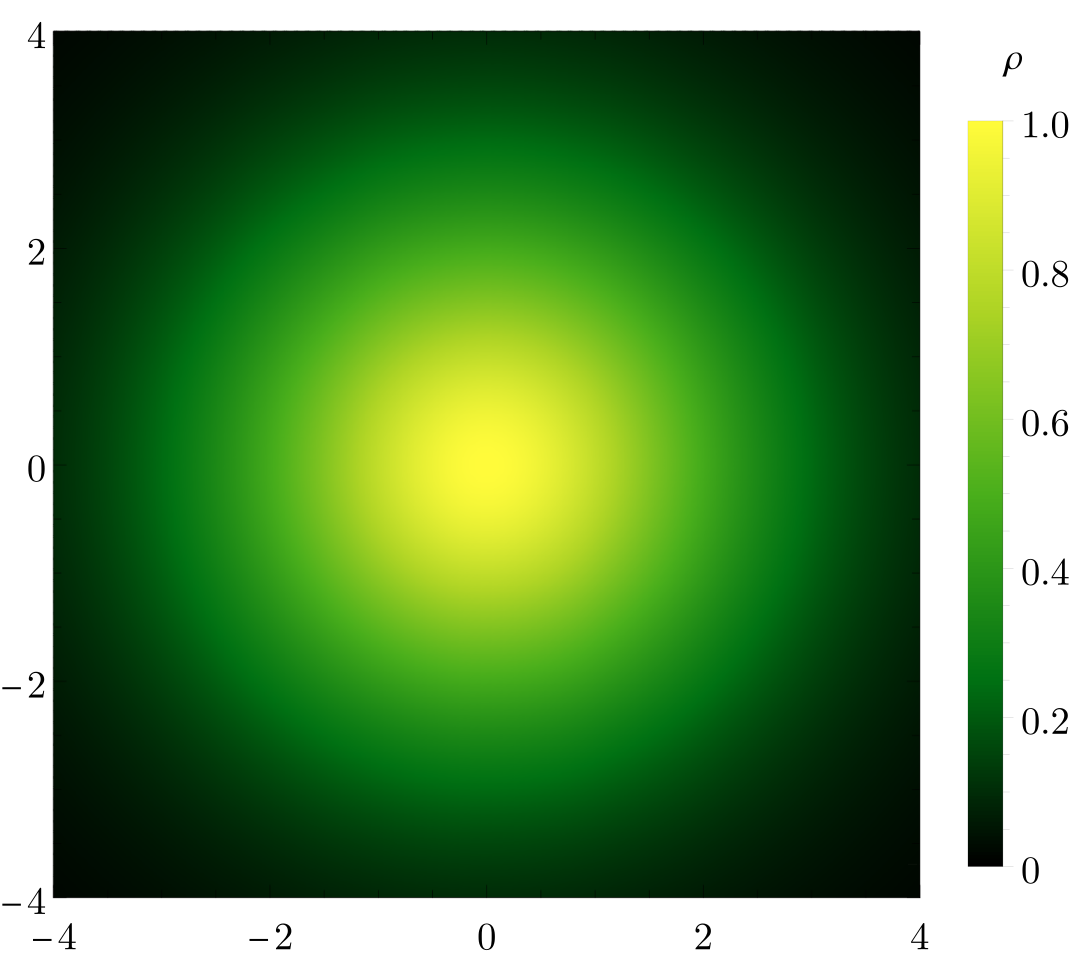}
   		\hfill
	 	\includegraphics[scale=0.4]{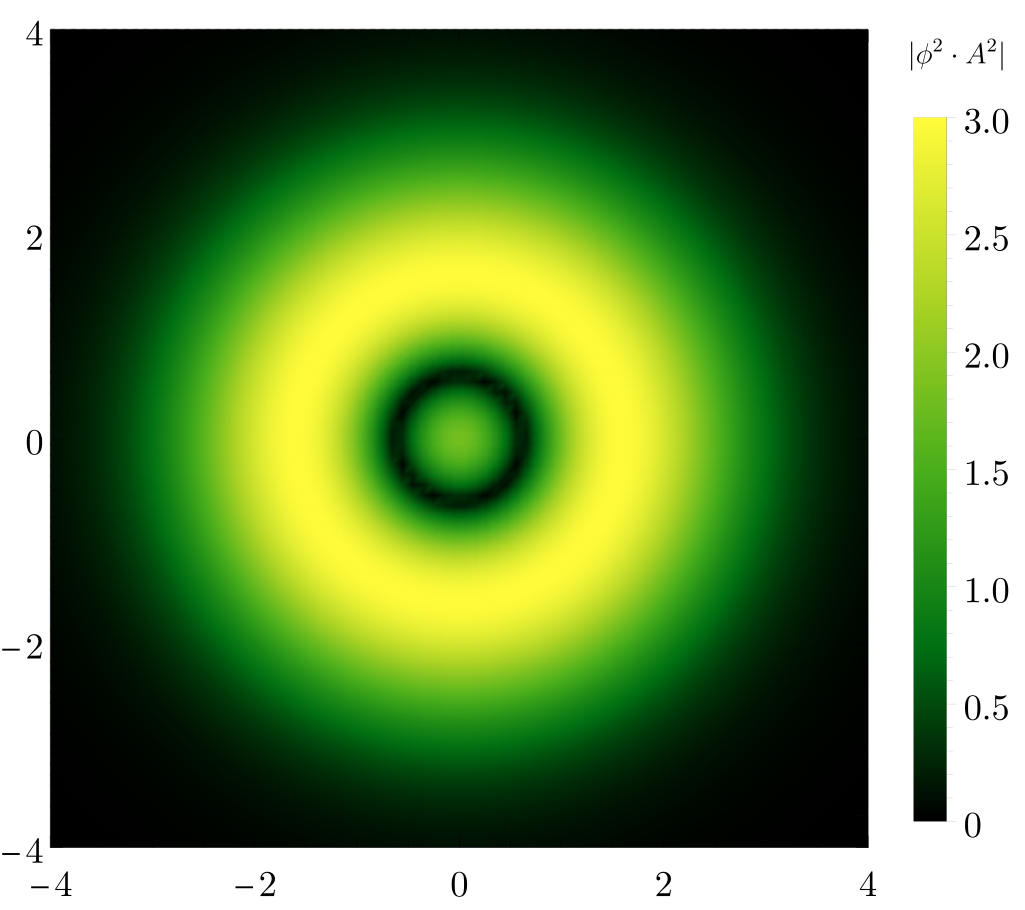}
	   	\caption{Minimally--coupled ($\alpha=0$) SPS with $\omega=0.758$ and $\gamma=0.834$ ($\mathcal{V}=0.824$), $M=0.885$, $Q=0.805$ (red circle in~\autoref{F3.2}). (Left) metric and matter functions radial profile: (solid black) metric function $\sigma$; (Solid yellow) matter function $m$; (Solid green) vector field function $f$; (Dot-dashed purple) energy density, $\rho$; (Dotted red) scalar field amplitude, $\phi$; And (dashed blue) vector field function $g$.(Middle) Energy density $\rho$ (normalized relative to the maximum). (Right) $|\phi^2\mathbf{A}^2|$ (min--max normalized). In both density plots, black (yellow) color represents the absence (maximum value) of $\rho$ or $|\phi^2\mathbf{A}^2|$. The solution position in the domain of existence is represented by a red circle in \autoref{F3.2}. Observe that the solution is regular everywhere.}
	   	\label{F3.1}
	\end{figure}

%
	When $\alpha=0$, the metric and matter functions exhibit a behavior akin to that of the corresponding single--field BSs -- see \autoref{F3.1} (left): $\phi$ and $f$ pile up around $r=0$ and have a global maximum there, whereas the minimum of $g$ is off-center. Also, $\phi$ and $g$ are nodeless, while $f$ has one node. Besides, most of the energy is concentrated around $r=0$ -- see~\autoref{F3.1} (middle). In addition, for future reference, \autoref{F3.1} (right) shows the shape of $|\Phi|^2\mathbf{A}^2$, which is proportional to the energy density contribution coming from the interaction term. 
	\begin{figure}[H]
		\centering
		\includegraphics[scale=0.80]{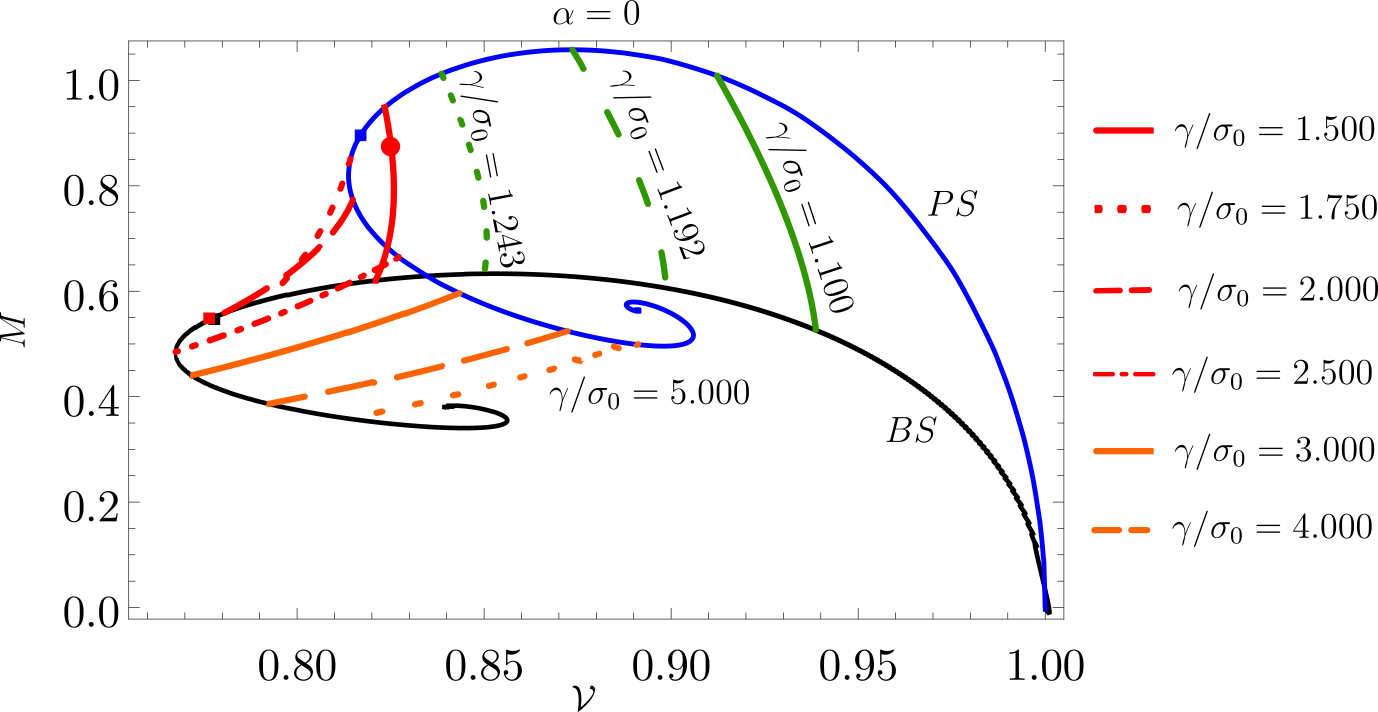}
		\hfill
   	    \includegraphics[scale=0.75]{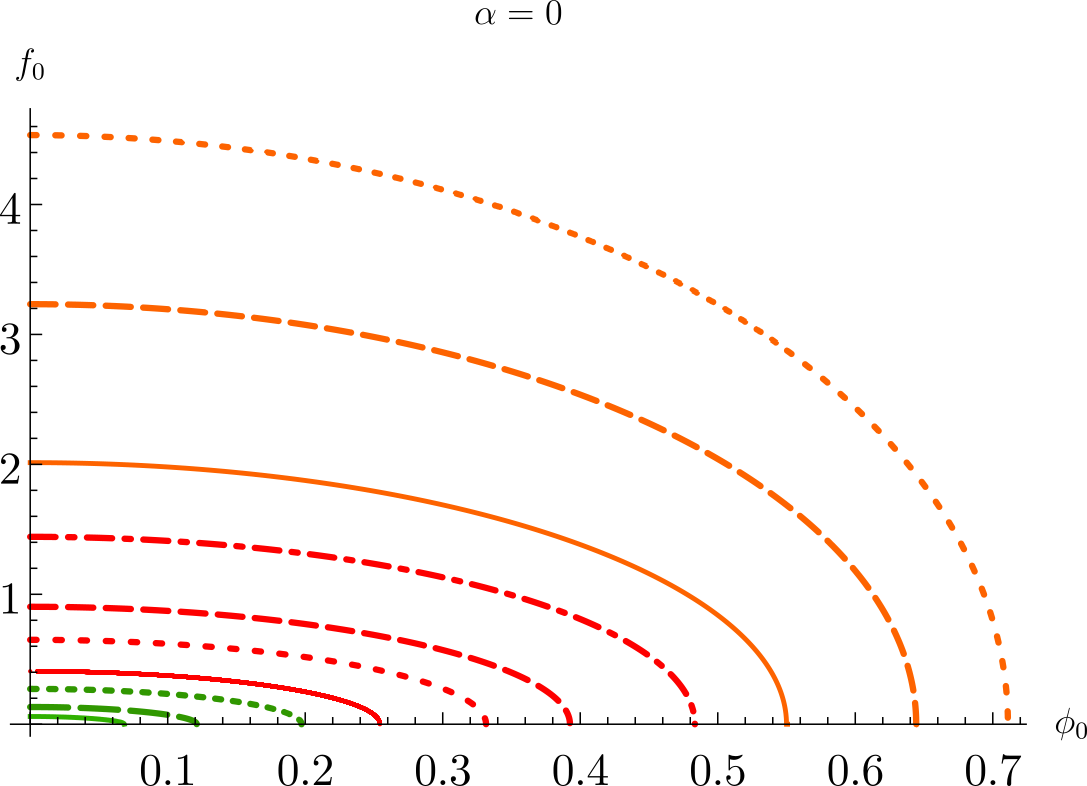}
	  	\caption{Domain of existence of minimally coupled ($\alpha=0$), non--synchronized $(\omega\neq\gamma)$ SPSs in a mass--effective frequency (left) and $f_0$--$\phi_0$ diagram (right) for: a solution with both ends in the stability region $\gamma /\sigma _0 =1.100$ (solid green); A solution that starts at the PS maximum (PS stability transition) $\gamma /\sigma _0 = 1.192$ (dashed green); A solution that starts in the unstable PS branch and ends in the SBS stability transition (maximum SBS mass) $\gamma /\sigma _0 =1.243$ (dotted green); An SPS line that has both ends pertubatively unstable but are energetically stable $\gamma /\sigma _0 =1.500$ (solid red); An SPS line that has both ends perturbatively stable but the correspondent SBS is energetically stable $\gamma /\sigma _0 =1.7500$;  And a set of configurations that have both pure configurations simultaneously perturbatively and energetically unstable: $\gamma /\sigma _0 = 2.000$ (dashed red), $\gamma /\sigma _0 = 2.500$ (dot-dashed red), $\gamma /\sigma _0 = 3.000$ (solid orange), $\gamma /\sigma _0 = 4.000$ (dashed orange), $\gamma /\sigma _0 = 5.000$ (dotted orange). The red circle corresponds to the minimally--coupled SPS in~\autoref{F3.1}. The squares denote solutions with $M=Q$, with those to the right (left) having $M<Q$ ($M>Q$) and thus positive (negative) binding energy. Observe that the SPS solutions continously connect the pure PS configuration to the pure SBS configuration.}
	 	\label{F3.2}
	\end{figure}

	When $\omega\neq\gamma$, the space of solutions is spanned by $\{\omega,\gamma,M\}$, thus being 3--dimensional. However, since they connect single--field BSs, it is easier to visualize it in terms of the \textit{effective frequency}\footnote{The effective frequency $\mathcal{V}$ does not have physical meaning, since the scalar (vector) field oscillates at it's own frequency $\omega$ ($\gamma$). This new quantity is here introduced to ease data visualization, and make clear that SPSs connect SBSs to PSs.}
	\begin{align}
       \mathcal{V}=\frac{\omega Q_\Phi+\gamma Q_A}{Q_\Phi+Q_A}\ .
	\end{align}	
	Pure SBSs (PSs) are recovered when $Q_A=0$ ($Q_\Phi=0$), hence $\mathcal{V}=\omega$ ($\mathcal{V}=\gamma$).

	Inspection of~\autoref{F3.2} (left) reveals interesting features. Firstly, the mass of a minimally--coupled SPS ranges between that of a SBS with frequency $\omega$ and that of a PS with frequency $\gamma$, in agreement with the values $\phi_0$ and $f_0$ take along the existence line of SPSs -- see~\autoref{F3.2} (right). Despite the minimal interaction between the fields (which is purely gravitational), these stars cannot be regarded as linear combinations of the corresponding pure BSs, as their mass is always smaller than the corresponding PSs. This picture can change when the fields are non--minimally coupled (to each other) and/or synchronized -- see~\autoref{sec:3.2}. Secondly, there can be different solutions with the same mass and effective frequency. This is clear for $\gamma/\sigma_0\in\{1.5,2.5\}$. 
	
	Thirdly, it is well--known that single--field BSs are stable against linear radial perturbations for frequencies greater that the one that maximizes their mass ~\cite{Gleiser:1988ih,Brito:2015pxa}. \autoref{F3.2} (left) shows that there are families connecting $(i)$ stable SBSs to stable PSs; $(ii)$ unstable SBSs to unstable PSs; and $(iii)$ stable SBSs to unstable PSs. While for the two former families $(i)$--$(ii)$ it is plausible that the connecting SPS solutions are always stable/unstable, the same cannot be true for the latter case. In general, it is not evident how $(iii)$ SPSs would behave when radially perturbed. However, if solutions sufficiently close to the branch out points retain the stability of the single--field BS, then one expects stable solutions that originate from a stable SBS to become unstable at a certain point close to the corresponding unstable PS. Such behavior is common to every $\alpha$ value studied.
	
	Finally, fixing $M$, the total Noether charge $Q=Q_\Phi+Q_A$ increases as $\gamma/\sigma_0$ decreases. For sufficiently large mass, $Q>M$ for $\gamma/\sigma_0\in\{1.100,1.192,1.243,1.500\}$ (positive binding energy), and $Q<M$ for $\gamma/\sigma_0\in\{1.75,2.00,2.50,3.00,4.00,5.00\}$ (negative binding energy). 
%
	\subsection{Non--minimal coupling ($\alpha\neq0$)}\label{sec:3.2}

	The addition of a non--minimal coupling between the fields provides the theory with a new interaction and associated possibilities. Its nature (either attractive or repulsive) depends on $\text{sgn}(\alpha\mathbf{A}^2)$ and can substantially change the physical properties of SPSs. The interaction is said to be attractive (repulsive) when $\text{sgn}(\alpha\mathbf{A}^2)=+1$ $(-1)$, since the last term in \eqref{eq:2.1} has the same (opposite) sign as that of the mass terms. In general, $\mathbf{A}^2$ does not have a fixed sign. This means that the nature of the interaction can change throughout the spacetime. By continuity, for sufficiently small positive (negative) values of $\alpha$, the interaction is repulsive (attractive) close to the origin, and becomes atractive (repulsive) at $r\sim\mu$ towards infinity.
	
		\subsubsection{Positive coupling ($\alpha>0$)}\label{sec:3.2.1}

	When $\alpha>0$, both fields can either oscillate in synchrony ($\omega=\gamma$) or not ($\omega\neq\gamma$). It was only possible to find numerically synchronized configurations for sufficiently high values of the interaction coupling constant, $\alpha\gtrsim 70$, while non--synchronized configurations were found for a wider range of $\alpha$. Let us start with the most generic case ($\omega\neq\gamma$).

		\subsubsection*{Non--synchronized configurations ($\omega\neq\gamma$)}\label{sec:3.2.2}

\begin{figure}[H]
		\centering
        \includegraphics[scale=0.75]{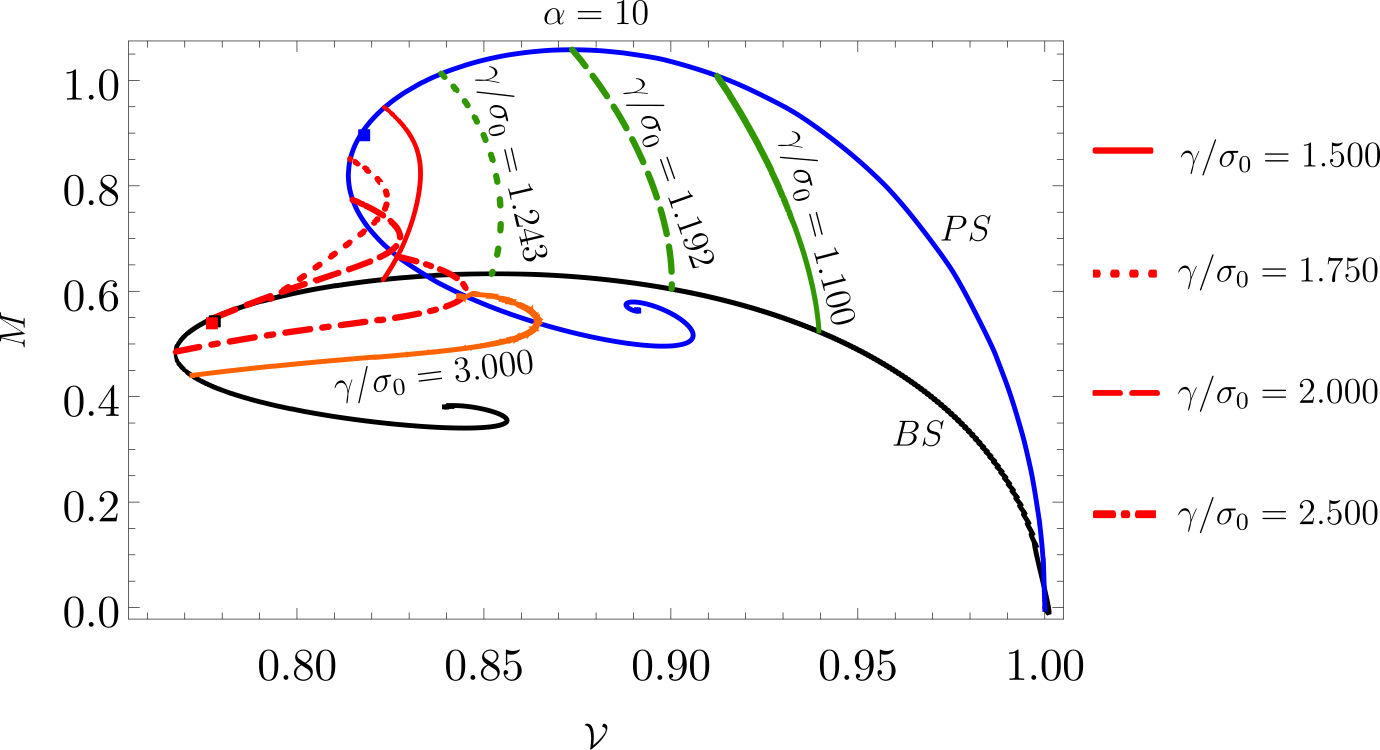}\hfill
        \includegraphics[scale=0.7]{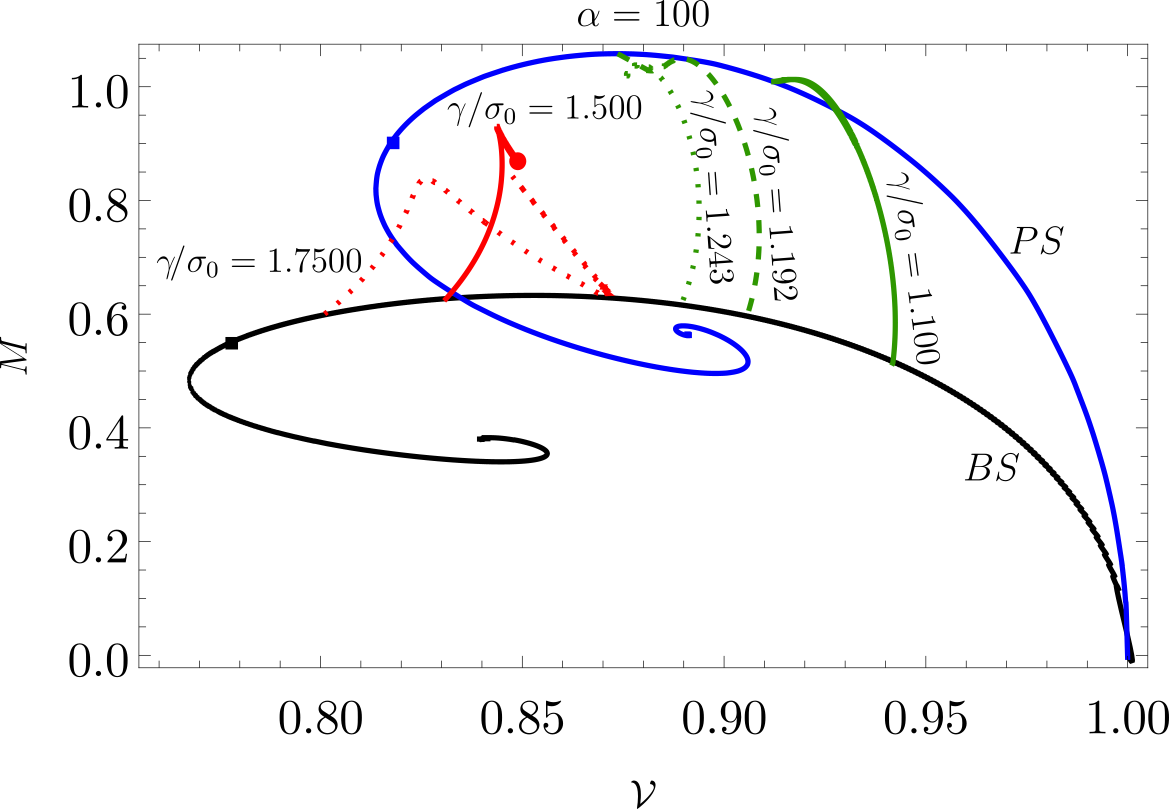}
       	\caption{(Top) domain of existence of non--minimally coupled ($\alpha\neq0$), non--synchronized ($\omega\neq\gamma$) SPSs with $\alpha=10$ (left) and $\alpha = 100$ (right) in a mass--effective frequency diagram for: a solution with both ends in the stability region $\gamma /\sigma _0 =1.100$ (solid green); A solution that starts at the PS maximum (PS stability transition) $\gamma /\sigma _0 = 1.192$ (dashed green); A solution that starts in the unstable PS branch and ends in the SBS stability transition (maximum SBS mass) $\gamma /\sigma _0 =1.243$ (dotted green); An SPS line that has both ends pertubatively unstable but are energetically stable $\gamma /\sigma _0 =1.500$ (solid red); An SPS line that has both ends perturbatively stable but the correspondent SBS is energetically stable $\gamma /\sigma _0 =1.7500$;  And a set of configurations that have both pure configurations simultaneously perturbatively and energetically unstable: $\gamma /\sigma _0 = 2.000$ (dashed red), $\gamma /\sigma _0 = 2.500$ (dot-dashed red), $\gamma /\sigma _0 = 3.000$ (solid orange). The red dot represents the last (numerically) obtainable solution with $\alpha=100$ and $\gamma/\omega_0=1.500$.}
	 	\label{fig:3.3}
	\end{figure}

	Non--synchronized SPSs with a relatively small $\alpha>0$ bear a very close resemblance to minimally coupled SPSs (see~\autoref{sec:3.1}), as shown in~\autoref{fig:3.3} (top) for $\alpha\in\{10,100\}$ (smaller values of $\alpha$ possess a neglegible difference with the minimally coupled case -- see~\autoref{F3.2}). Fixing $\gamma/\sigma_0$, the higher the coupling constant, the higher the value $\phi_0$ takes as approaching the existence line of SBSs (\textit{i.e.} as $f_0\rightarrow0$).
Moderate values of $\alpha$ do not impact the shape of the lines nor the maximum mass of SPSs (which is still equal to the mass of the PS they branch out from). However, when $\alpha \sim 100$, the line wiggles close to the branch out point, generating new branches of solutions, some of which are heavier than the corresponding PS -- see \autoref{fig:3.3} (right) for $\alpha =100$.
 
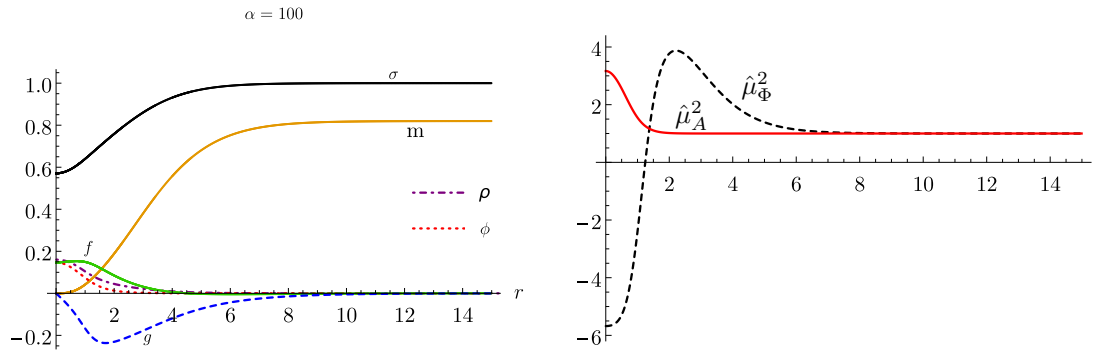
\begin{figure}[H]
			\centering
     	      \begin{picture}(0,0)
		  	 \put(280,98){$\hat{\mu} ^2 _\Phi$}
             \put(255,87){$\hat{\mu} ^2 _A$}
	   	  \end{picture}
        \includegraphics[scale=0.75]{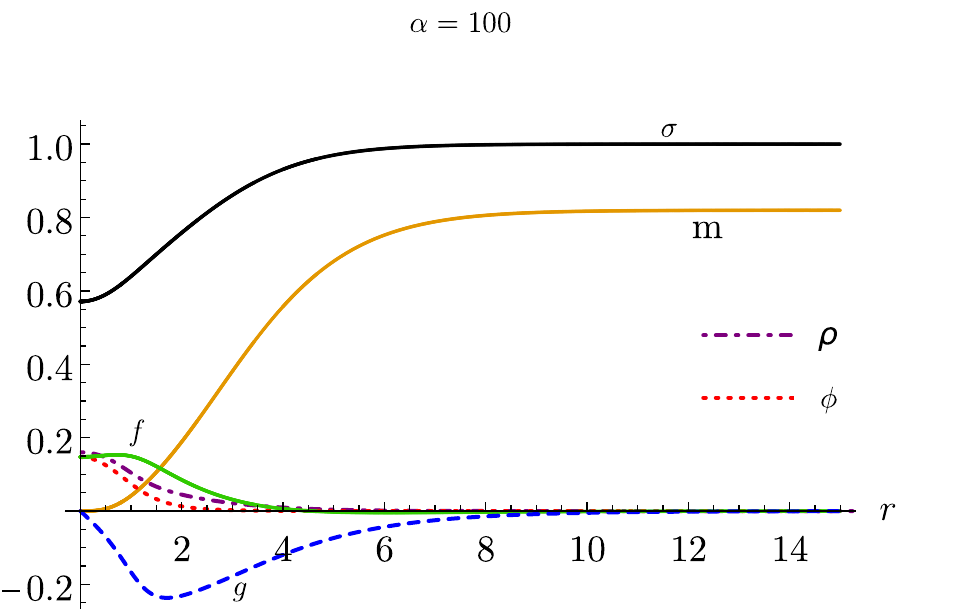}
        \includegraphics[scale=0.75]{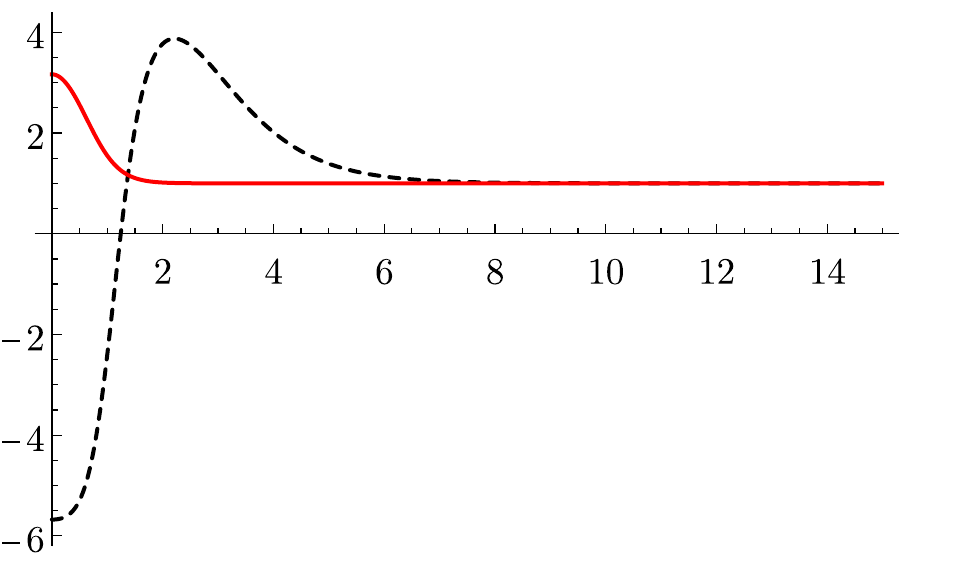}
	   	\caption{Non--minimally coupled, synchronized SPS with $\mathcal{V}=0.857$ ($\gamma/\sigma _0=1.500$), $\alpha =100$, $M=0.812$ and $Q=0.811$.(Left) metric and matter functions radial profile: (solid black) metric function $\sigma$; (Solid yellow) matter function $m$; (Solid green) vector field function $f$; (Dot-dashed purple) energy density, $\rho$; (Dotted red) scalar field amplitude, $\phi$; And (dashed blue) vector field function $g$. (Right) square of the effective scalar and vector masses for the last (numerically) obtainable solution with $\alpha=100$ and $\gamma/\omega_0=1.500$ -- denoted by a red dot in \autoref{fig:3.3} (right). (Right) scalar (dashed black) and vector (solid red) field's effective mass as a function of the radial coordinate for the previous solution. The effective mass of the scalar field becomes imaginary close to the origin. Observe that the solution is everywhere regular.}
	 	\label{fig:3.3a}
	\end{figure}

	 For sufficiently high values of $\gamma/\sigma_0$, it was not possible to obtain solutions continuously connecting single--field BSs. For $\gamma/\sigma_0=1.500$, the last (numerically) obtainable solution has $\hat{\mu}_\Phi^2<0$, as shown in~\autoref{fig:3.3} (bottom) -- denoted by a red circle in ~\autoref{fig:3.3a}.
	Finally, while for $\alpha =10$ the stability argument put forward in \autoref{sec:3.1} seems apply, an \textit{addendum} has to be made for $\alpha = 100$. From catastrophy theory arguments \cite{Kleihaus:2011sx,Kusmartsev:1990cr,Tamaki:2010zz,Tamaki:2011zza}, one could argue that for each turning point in the domain of existence $\{ \mathcal{V},M\}$, like the one shown in \autoref{fig:3.3} (right) for $\gamma/\sigma_0=\{ 1.100,1.192,1.243\}$, there is a transition in the stability of the solutions. The latter argument seems to agree with the observed results, however $\mathcal{V}$ is not a proper frequency and can give erroneous results. Further studies have to be performed, but the data seems to indicate the existence of sections of the SPS line with opposing stability to the rest.
%
		\subsubsection*{Synchronized configurations ($\omega=\gamma$)}\label{sec:3.2.3}

\begin{figure}[H]
	 	 \centering
	 	 \includegraphics[scale=0.75]{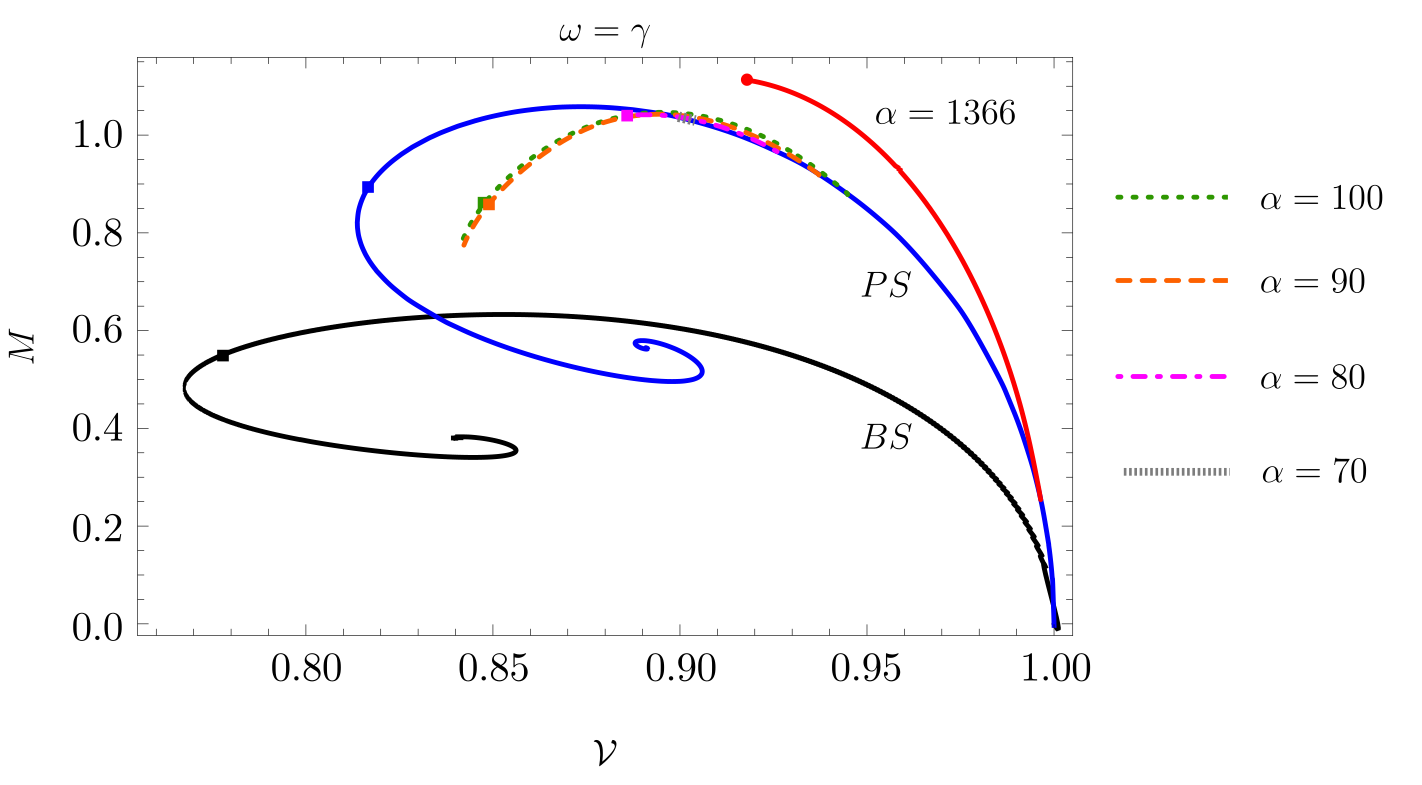}
	 	 \includegraphics[scale=0.75]{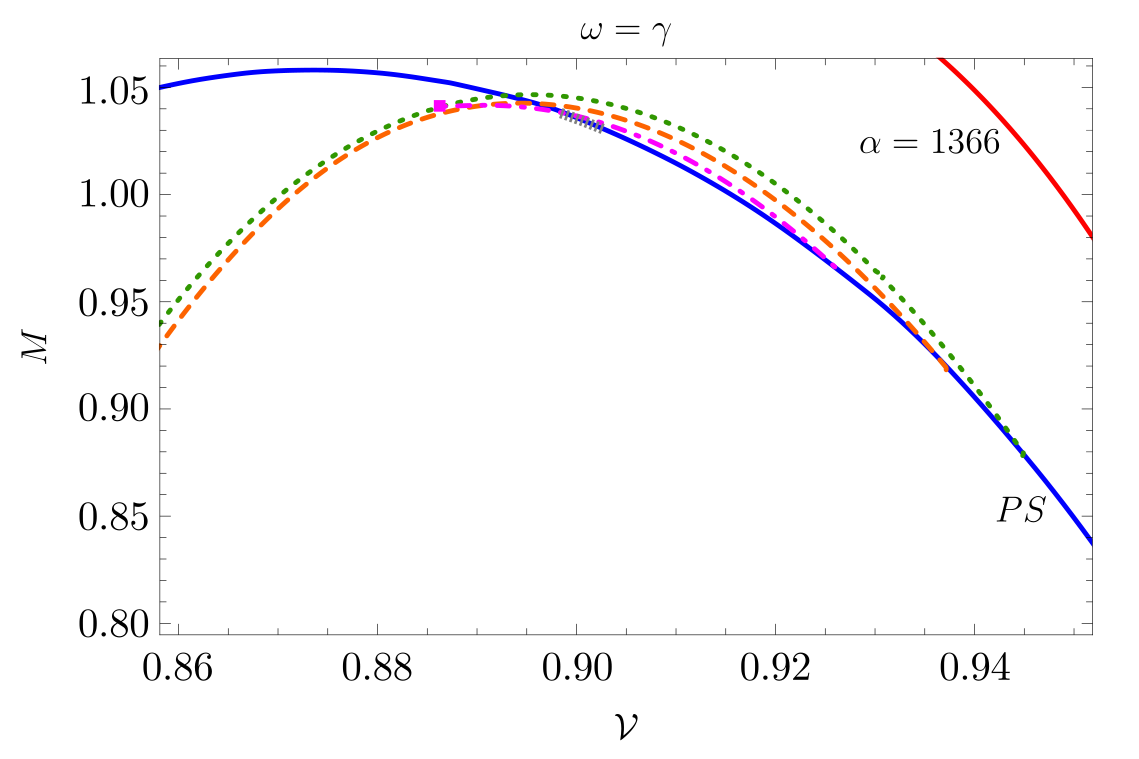}
	 	 \caption{(Left) domain of existence of non--minimally coupled ($\alpha=0$), synchronized ($\omega=\gamma$) SPSs in a mass--effective frequency diagram, for: $\alpha = 1366$ (solid red) maximum value numerically obtainable; $\alpha = 100$ (dotted green); $\alpha = 90$ (dashed orange); $\alpha = 80$ (dot-dashed violet) and; $\alpha=70$ (dotted grey) close to the minimum value of $\alpha$ for which synchronized solutions exist. (Right) zoom-in of the region where the majority of the lines branch out from the PS. Note that for high enough values of $\alpha$ one can obtain SPS configurations that have an higher mass than the pure PS configuration, from which all lines seem to emerge from.}
		 \label{fig:3.4}
	\end{figure}

	When the fields are synchronized, $\mathcal{V}=\omega=\gamma$,  and the space of solutions is spanned by $\{\mathcal{V},M\}$. The existence of such configurations turns out to be very sensitive to the coupling constant, as shown in~\autoref{fig:3.4} for $\alpha\in\{70,80,90,100,1366\}$. The (finite) range of frequencies for which SPSs exist shrinks as $\alpha$ decreases and appears to vanish completely for $\alpha\lesssim70$ -- see also Table~\ref{T1} .
	
\begin{figure}[H]
		 \centering
		 \includegraphics[scale=0.6]{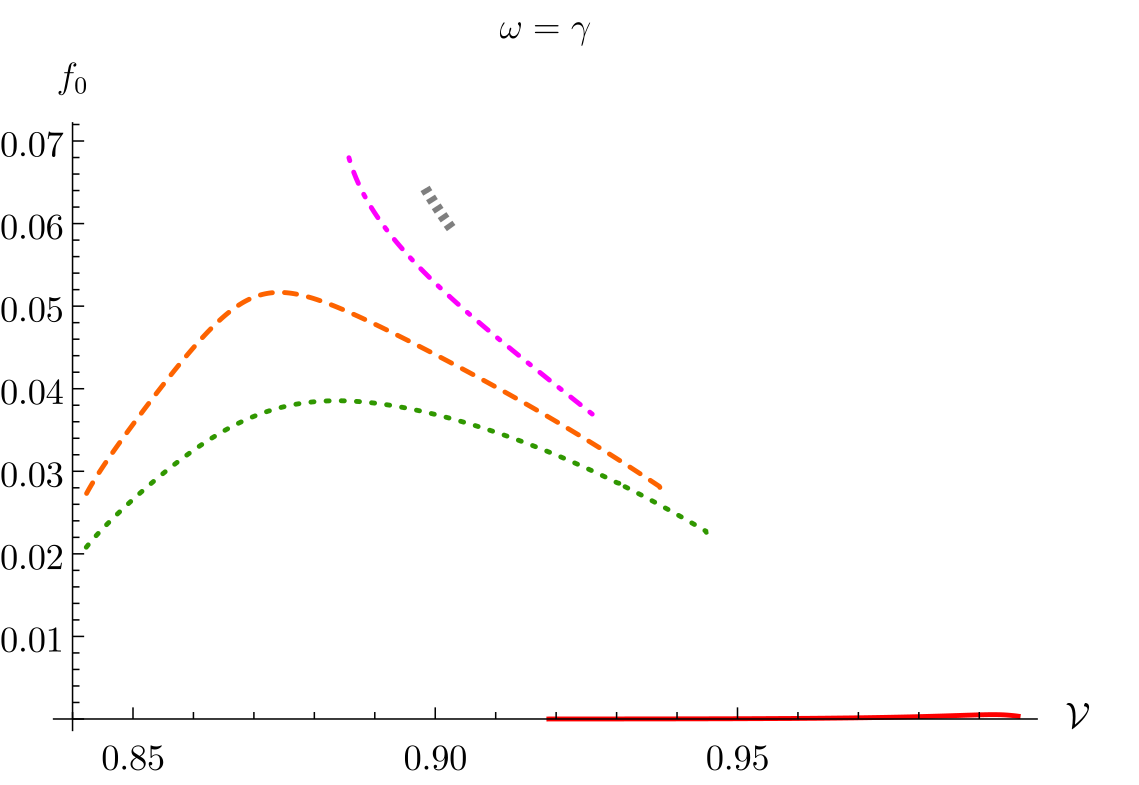}
         \includegraphics[scale=0.6]{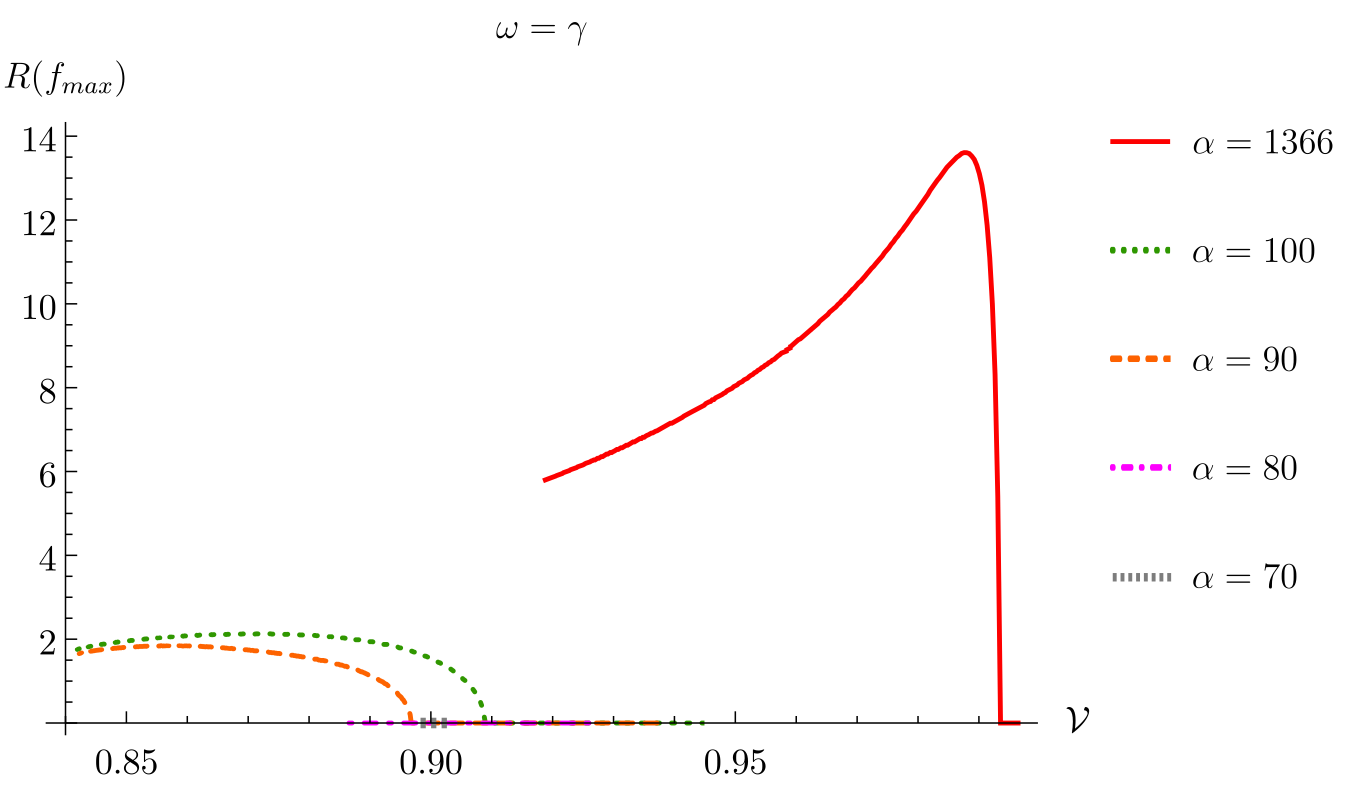}
	 	 \caption{Same as in~\autoref{fig:3.4}, but in a $f_0$--effective frequency diagram (left) and in a $R(f_\text{max})$--effective frequency diagram (right) for: $\alpha = 1366$ (solid red) maximum value numerically obtainable; $\alpha = 100$ (dotted green); $\alpha = 90$ (dashed orange); $\alpha = 80$ (dot-dashed violet) and; $\alpha=70$ (dotted grey) close to the minimum value of $\alpha$ for which synchronized solutions exist. Note that $f$ has its maximum at the origin for $\alpha\in\{70,80\}$. The decrease in $f_0$ seems to be associated with the end of synchronized set of configurations shown in \autoref{fig:3.4}.}
		 \label{fig:3.5}
	\end{figure}

	 SPSs branch out from PSs with increasing frequency as $\alpha$ increases\footnote{Since the neighborhood of $(\omega,M)=(\mu,0)$ is hard to probe, it is not clear whether the frequency at the branch out approaches $\mu$ when $\alpha$ tends to infinity or a finite value.}. This seems intuitive: a more dilute field requires a stronger interaction to yield the same configuration. This is in agreement with the trend in the mass, which is greater for larger values of $\alpha$ (when comparable) and can even be greater than that of a PS with the same frequency. Although the solutions branch out from the PS line, they do not join the SBS line\footnote{No evidence for synchronized configurations branch out from the SBS line was found.}, as opposed to non--synchronized configurations. 
	For $\alpha\in\{70,80\}$, the mass increases monotonically as one moves away from the branch out point (\textit{i.e.} as $\mathcal{V}$ decreases). However, for $\alpha\in\{90,100\}$, it reaches a maximum and then decreases. While $f_0$ follows a similar trend -- see~\autoref{fig:3.5} (left) --, $\phi_0$ increases as $\mathcal{V}$ decreases regardless of the value $\alpha$ takes (not shown).
	\begin{figure}[H]
	 	 \centering
	 	 \includegraphics[scale=0.51]{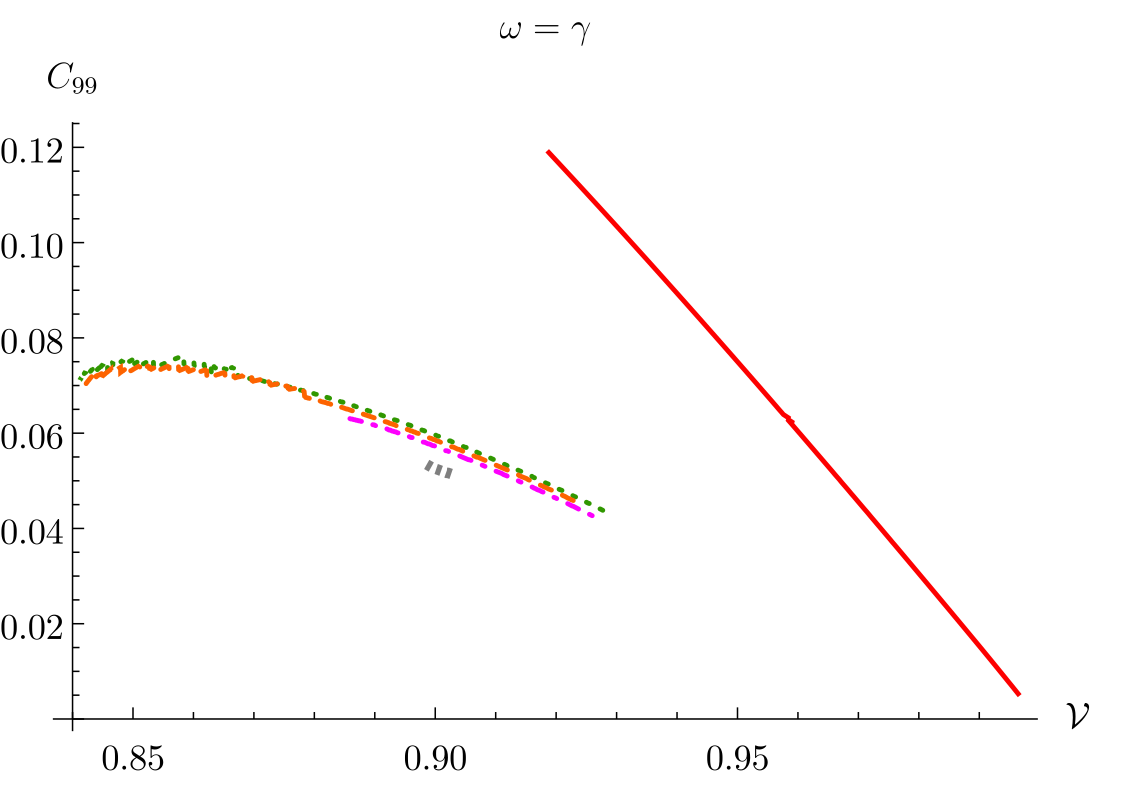}
	 	 \hfill
       	 \includegraphics[scale=0.51]{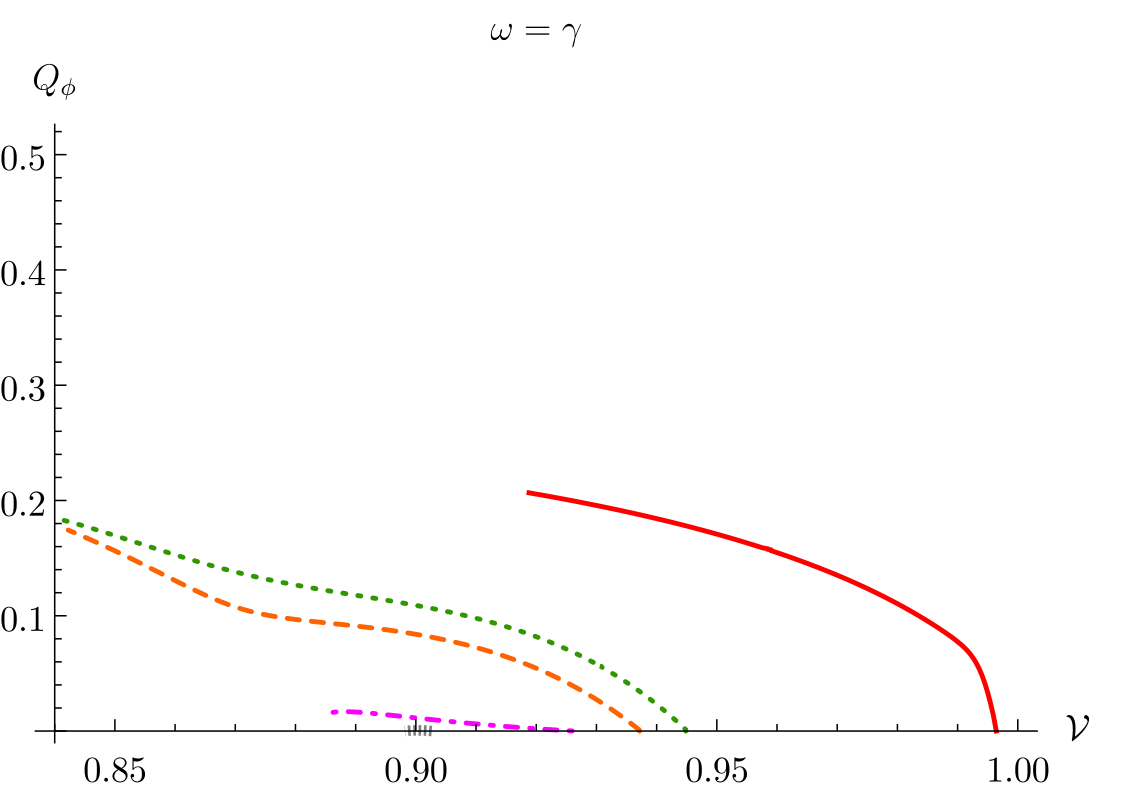}
       	 \hfill
      	 \includegraphics[scale=0.51]{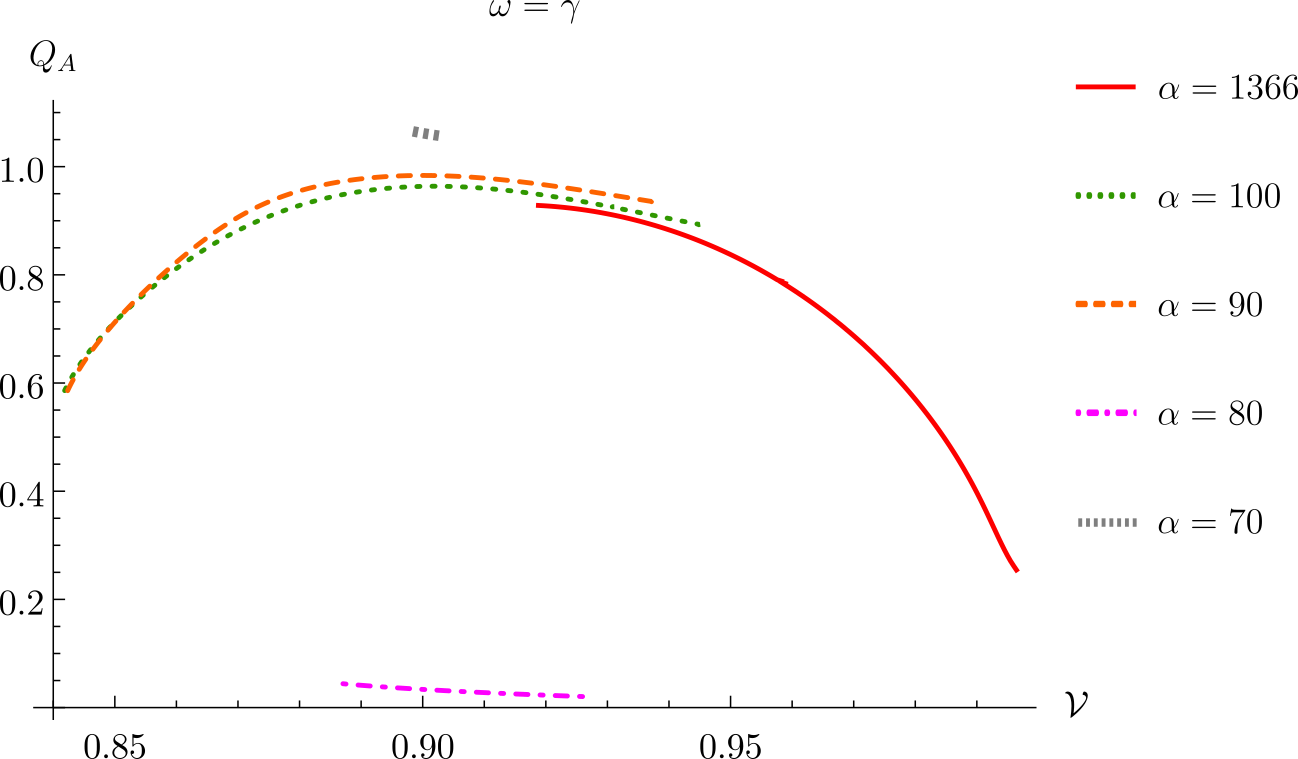}
	 	 \caption{Graphical representation of several characteristic quantities of non--minimally coupled, synchronized SPSs in Fig.~\ref{fig:3.4}: compactness (left), scalar (middle) and vector (right) Noether charges for: $\alpha = 1366$ (solid red) maximum value numerically obtainable; $\alpha = 100$ (dotted green); $\alpha = 90$ (dashed orange); $\alpha = 80$ (dot-dashed violet) and; $\alpha=70$ (dotted grey) close to the minimum value of $\alpha$ for which synchronized solutions exist.}
		 \label{fig:3.6}
	\end{figure}
	\autoref{fig:3.6} reveals the effects of changing $\alpha$ on the compactness and (scalar and vector) Noether charges of the solitonic stars. The compactness, defined as twice $99\%$ of the star's mass divided by the perimetral radius that contains it ($C_{99}=2 M /R_{99}$), increases with increasing $\alpha$. Nevertheless, the stars are always dilute and non--relativistic when $\mathcal{V}\rightarrow\mu$. \autoref{fig:3.6} also shows that the scalar (vector) Noether charge grows (drops) as $\alpha$ rises.
	
\begin{figure}[H]
    		 \centering
		 \includegraphics[scale=0.75]{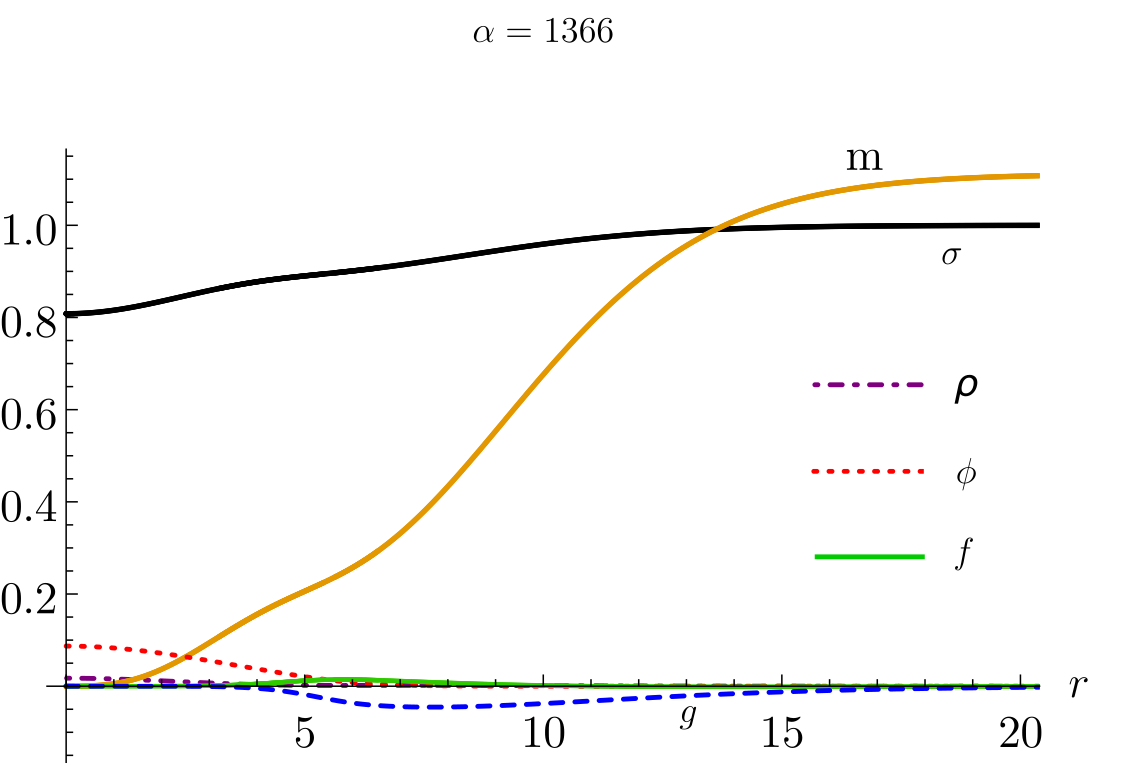}
	 	 \includegraphics[scale=0.45]{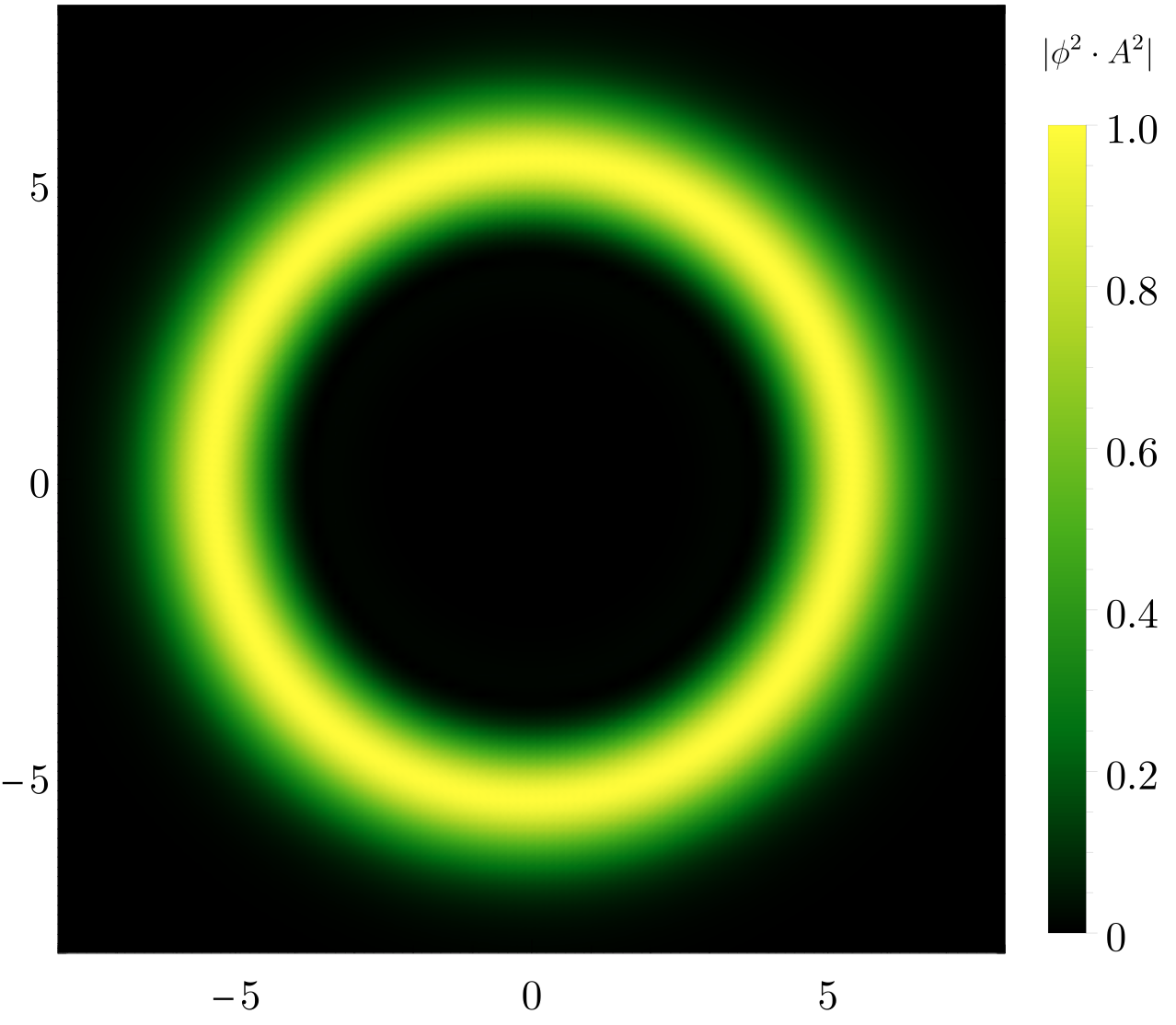}
	   	 \caption{Non--minimally coupled, synchronized SPS with $\mathcal{V}=0.758$ ($\gamma/\sigma _0=1.500$), $\alpha =1366$, $M=1.112$ and $Q=1.135$. (Left) metric and matter functions radial profile: (solid black) metric function $\sigma$; (Solid yellow) matter function $m$; (Solid green) vector field function $f$; (Dot-dashed purple) energy density, $\rho$; (Dotted red) scalar field amplitude, $\phi$; And (dashed blue) vector field function $g$. (Right) $|\phi^2\mathbf{A}^2|$ (min--max normalized), black (yellow) color represents the absence (maximum value) of $|\phi^2\mathbf{A}^2|$. Observe that the solution is regular everywhere.}
	 	 \label{fig:3.7}
		\end{figure}

	Also worth mentioning is the shift in the position of the maximum of $f$, $R(f_\text{max})$, which becomes off--center for sufficiently high values of the interaction coupling -- see \autoref{fig:3.5} (right). This creates a ring--like structure for the energy density contribution coming from the interaction term -- see~ \autoref{fig:3.7} (right). 
	Following the work done in \cite{Herdeiro:2021lwl}, one can also consider the presence of light rings and/or the inner edge of the accretion disk associated with a bound in the existence of a maximum of the angular velocity $\Omega$ along the orbits (see \cite{Herdeiro:2021lwl,olivares2020tell} for a more details). None of the solutions presented here possesses a light ring (in the case of pure BSs, the light rings only start to exist well inside the domain of existence spiral).
			\begin{figure}[H]
    		 \centering
		 \includegraphics[scale=0.8]{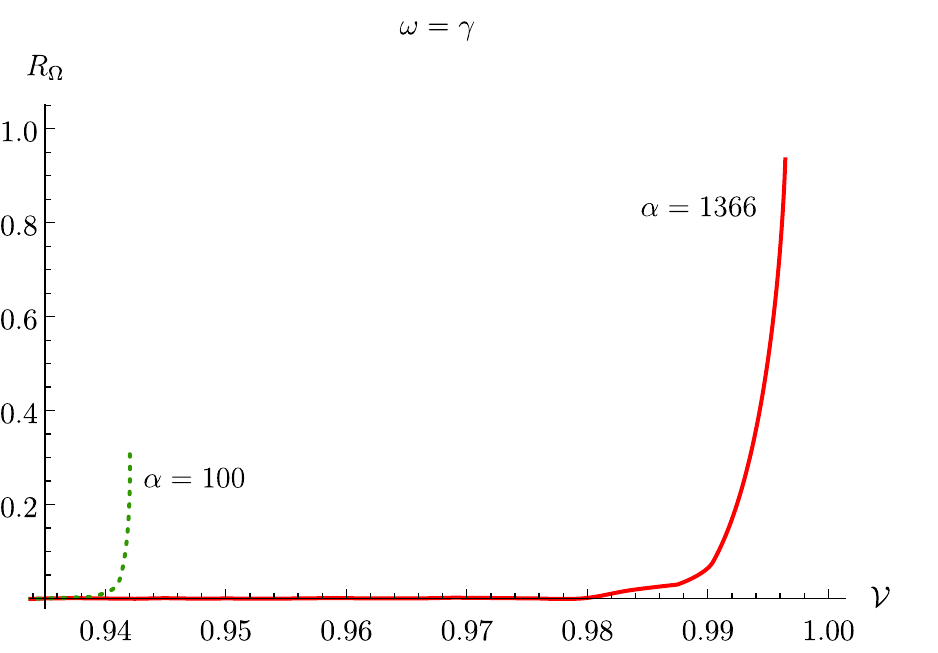}
	   	 \caption{Maximal value of the angular velocity radii, $R_\Omega$, of a particle around a boson star as a function of the effective frequency $\mathcal{V}$ for two values of the coupling interaction $\alpha =1366$ (solid red) and $\alpha = 100$ (dashed green).}
	 	 \label{fig:3.7a}
		\end{figure} 

	From previous works it is known that pure PSs possesses an inner edge of the accretion disk in the stable branch (while SBS do not). Our results show that SPS solutions, where both corresponding pure BS configurations do not have an inner edge of the accretion disk, also do not possess one (there are stable circular orbits all the way to the center of the BS); the same does not occur when the SPS configurations branches out from a region where the corresponding PS possesses an inner edge of the accretion disk but the SBS does not. In this case, there exist SPS solutions with an inner edge of the accretion disk close to the pure PS configuration, and as the SPS configurations tend to the SBS they lose the inner edge and gain stable circular orbits for any value of the radial coordinate.	
	
	As for the case of synchronized SPSs, these do not connect PSs to the SBS configurations, so the previous arguments can not be made. In the synchronized SPS case, while for $\alpha = \{ 70,80,90\}$ there are no solutions with an inner edge of the accretion disk -- the angular velocity has its maximum at the BS center -- the same does not occur for the higher coupling values -- see \autoref{fig:3.7a}. For $\alpha = \{ 100, 1366\}$, solutions start with a PS that contains an inner edge of the accretion disk outside the center $R_\Omega >0$ and this behavior occurs for the associated SPSs. However, as one goes away from the branch out point, one can observe that $R_\Omega \to 0$ and the SPS solutions lose the inner edge of the accretion disk.
\begin{table}[H]
 \centering
 \caption{Characteristic quantities of the branch out solution and the terminal point solutions for the non-minimally coupled synchronized ($\omega=\gamma=\mathcal{V}$) SPS solutions for five values of the interaction coupling $\alpha$.}\label{T1}
 \vspace{3mm}
\begin{tabular}{ c|cccc|cccc| }
	\multicolumn{1}{c}{ }& \multicolumn{4}{|c}{\bf Branch out point}&\multicolumn{4}{|c|}{\bf Terminal point} \\
\hline
    $\alpha$ & $\mathcal{V}$ & $M$ & $Q$ & $f_0$ & $\mathcal{V}$ & $M$ & $Q$ & $f_0$ \\
 \hline
    $1366$ & $0.9964$ & $0.2541$ & $0.2544$ & $0.0003$ & $0.9188$ & $1.1116$ & $1.1350$ & $5\times 10^{-5}$\\
    $100$ & $0.9448$ & $0.8782$ & $0.8926$ & $0.0226$ & $0.8425$ & $0.6432$ & $0.6005$ & $0.0152$\\
    $90$ & $0.9372$ & $0.9183$ & $0.9352$ & $0.0285$ & $0.8411$ & $0.7466$ & $0.7244$ & $0.0253$\\
    $80$ & $0.9260$ & $0.9663$ & $0.9780$ & $0.0369$ & $0.8859$ & $1.0041$ & $1.0653$ & $0.0686$\\
    $70$ & $0.9026$ & $1.0306$ & $1.0572$ & $0.0595$ & $0.8982$ & $1.0385$ & $1.0661$ & $0.0645$\\

\end{tabular}
\end{table}
%

    \subsubsection{Negative coupling ($\alpha<0$)}\label{sec:3.2.4}
%
	The direct interaction between the fields when $\alpha<0$ is expected to be repulsive at the star's core (at least for sufficiently small $|\alpha|$), thus counteracting their gravitational attraction. In principle there should be a critical (negative) value $\alpha_c$ for which these opposite effects counterbalance each other. 

\begin{figure}[H]
		\centering
        \includegraphics[scale=0.8]{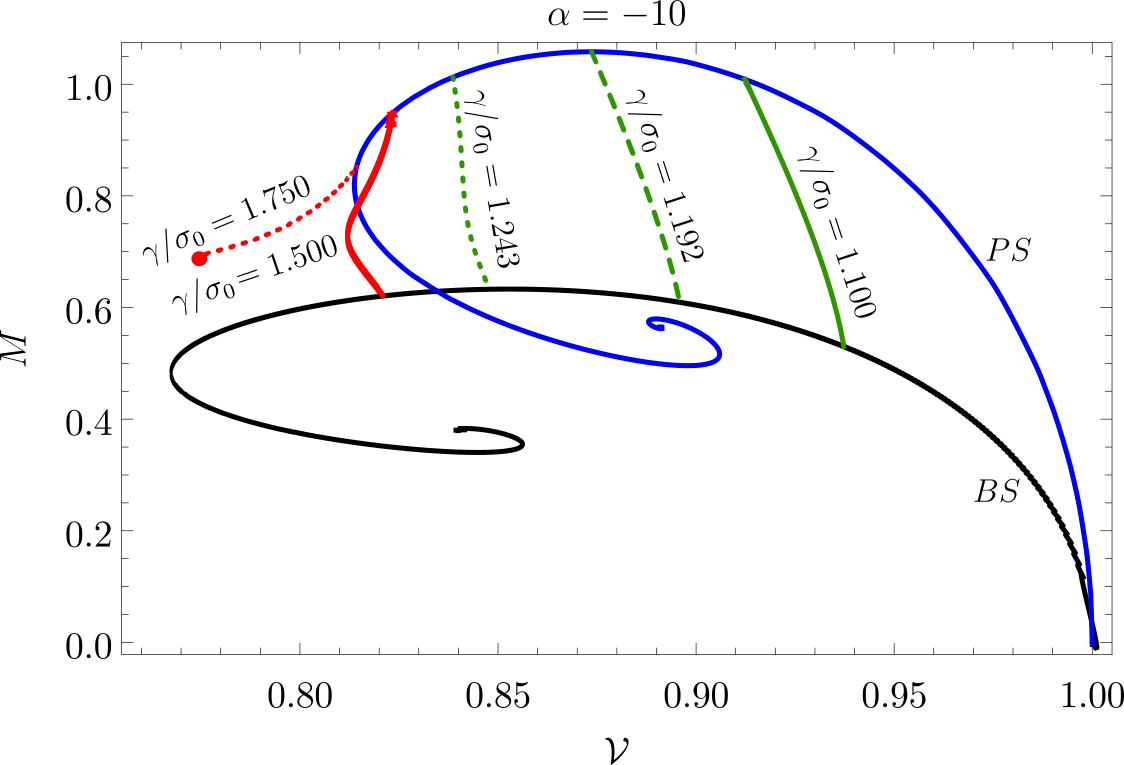}
	   	\caption{Domain of existence of non--minimally coupled, non--synchronized SPSs with $\alpha=-10$ in a mass--effective frequency diagram for: a solution with both ends in the stability region $\gamma /\sigma _0 =1.100$ (solid green); A solution that starts at the PS maximum (PS stability transition) $\gamma /\sigma _0 = 1.192$ (dashed green); A solution that starts in the unstable PS branch and ends in the SBS stability transition (maximum SBS mass) $\gamma /\sigma _0 =1.243$ (dotted green); A solution that starts in an energetical and pertubatively unstable PS and ends in an energetically stable SBS $\gamma /\sigma _0 =1.500$ (solid red); and a set of configurations that branch out from an unstable PS and do not end in an SBS configuration $\gamma /\sigma _0 = 1.750$ (dashed red). The red dot represents the last numerically obtainable solution for $\gamma /\sigma_0 = 1.750$ (solution profile graphically represented in \autoref{fig:3.8b}).}
	 	\label{fig:3.8a}
	\end{figure}

\begin{figure}[H]
		\centering
		\includegraphics[scale=0.75]{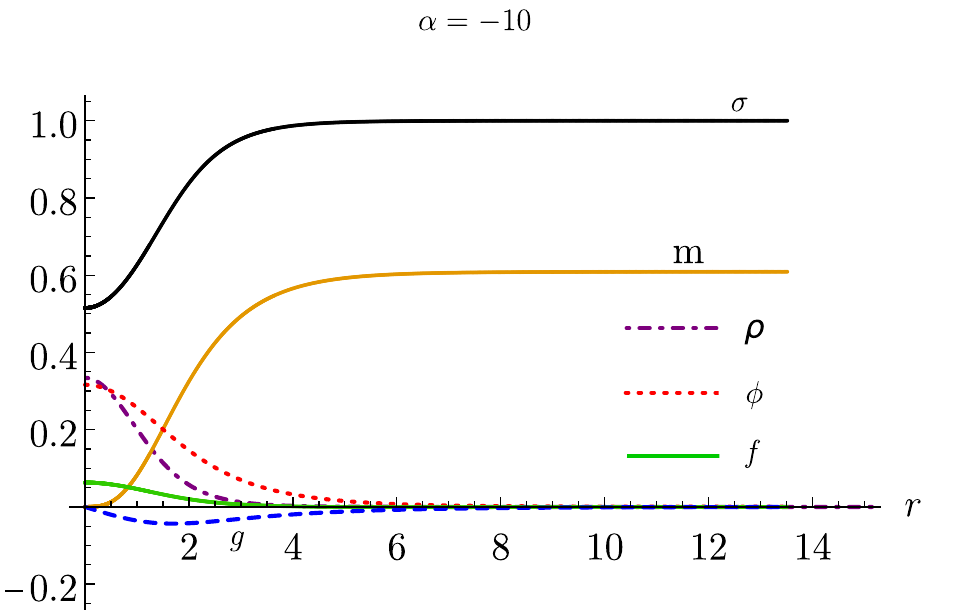}
		\qquad
	 	\includegraphics[scale=0.45]{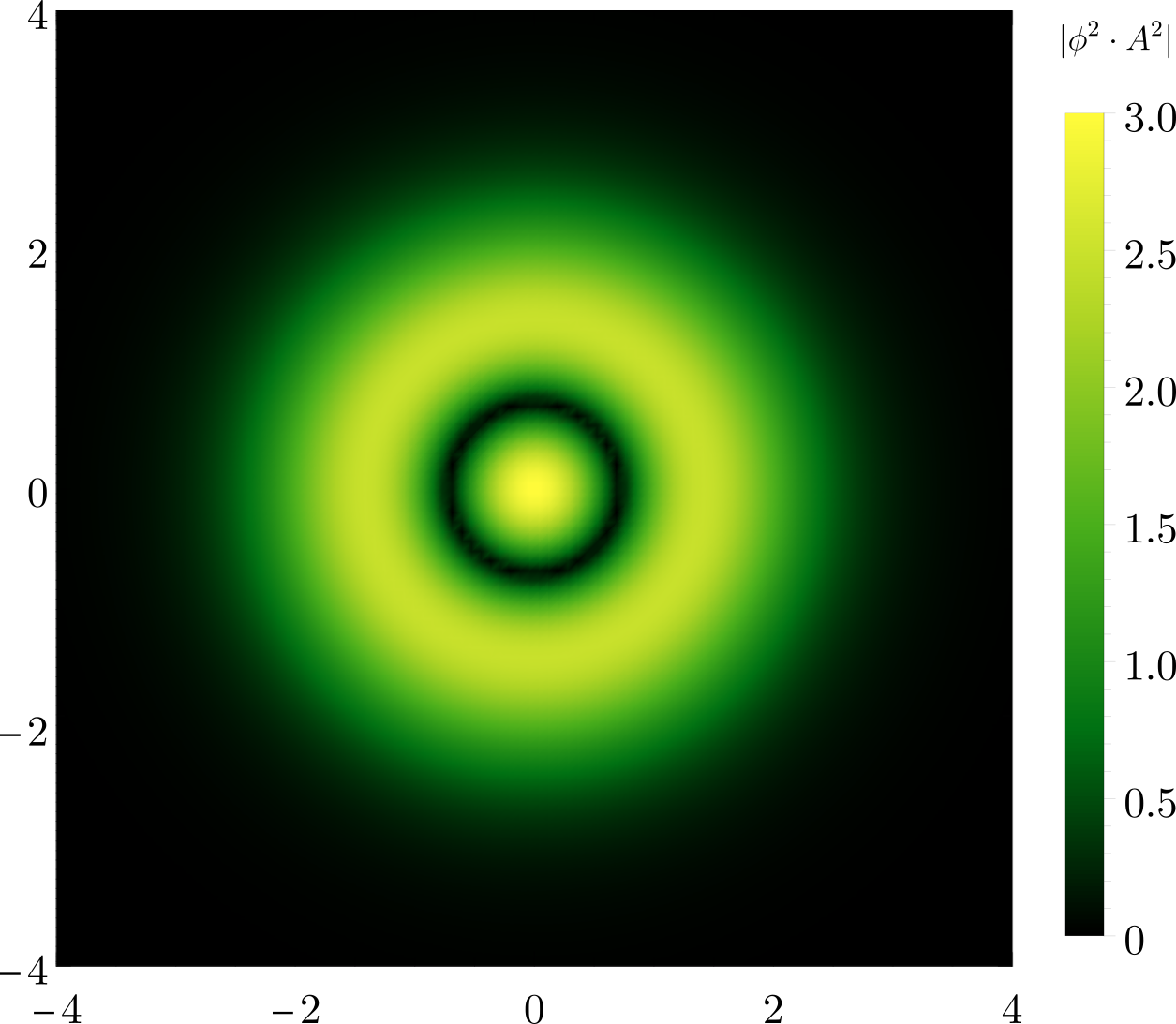}
	   	\caption{Non-minimally--coupled ($\alpha=-10$) SPS with $\omega=0.788$ and $\gamma=0.901$ ($\mathcal{V}=0.771$), $M=0.639$, $Q=0.621$ and $\gamma / \sigma_0=1.750$ (red circle in~\autoref{fig:3.8a}). (Left) metric and matter functions radial profile: (solid black) metric function $\sigma$; (Solid yellow) matter function $m$; (Solid green) vector field function $f$; (Dot-dashed purple) energy density, $\rho$; (Dotted red) scalar field amplitude, $\phi$; And (dashed blue) vector field function $g$. (Right) $|\phi^2\mathbf{A}^2|$ (min--max normalized), black (yellow) color represents the absence (maximum value) of $|\phi^2\mathbf{A}^2|$. Observe that the solution is regular everywhere.}
	 	\label{fig:3.8b}
	\end{figure}

	When $\alpha<\alpha_c$, repulsion dominates and prevents SPSs from forming. The critical value should depend on $\omega$ and $\gamma$, thus being hard to determine. For non--synchronized configurations, the domain of existence is  very similar to that of SPSs with $\alpha>0$, as shown in~\autoref{fig:3.8b} for $\alpha=-10$. Fixing $\gamma/\sigma_0$, the line moves towards higher effective frequencies as $|\alpha|$ grows, scaling down the minimum mass of SPSs. Despite the similitude, the radial profile of the matter functions may differ considerably from those of SPSs with $\alpha>0$, whereas $\mathbf{A}^2$ has an additional peak, also close to $r=0$, as one can infer from~\autoref{fig:3.8a}.
	 At last, we would like to point out that, while in the case of $\alpha >0$, high $\gamma /\sigma _0$ solutions exist in the vicinity of the SBS line but not in the vicinity of the PS line, for $\alpha <0$ the opposite occurs. In fact, in both cases, when attempting to construct SPS solutions close to the pure BS lines one observed that the pure configuration is unable to sustain any significant quantity of the other field. Which seems to indicate the existence of SPS lines that start in one of the pure configurations but do not end in the other.
%
		\subsubsection{Changing the coupling}\label{sec:3.2.5}
		\label{S3.4}
\begin{figure}[H]
			 \centering
			 \includegraphics[scale=0.85]{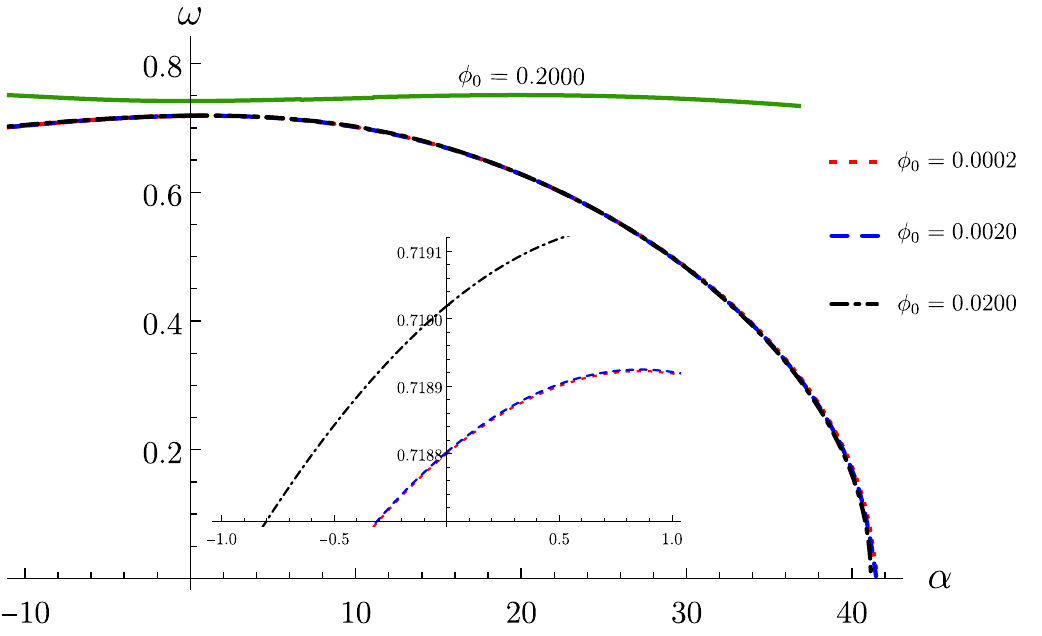}
	   		 \caption{Evolution of the scalar field frequency as a function of the interaction coupling for a set of (non)-minimally coupled SPS solutions with fixed initial scalar field amplitude: $\phi _0=0.0002$ (dot-dashed black); $\phi _0=0.0020$ (dashed blue); $\phi _0=0.0200$ (dotted red); and $\phi _0=0.2000$ (solid green). Observe that, while for positive values of $\alpha$ we manage to achieve $\omega \to 0$, no such limit was possible to obtain for $\alpha <0$.}
	 		 \label{fig:3.9}
			\end{figure}

\begin{figure}[H]
		\centering
		\includegraphics[scale=0.75]{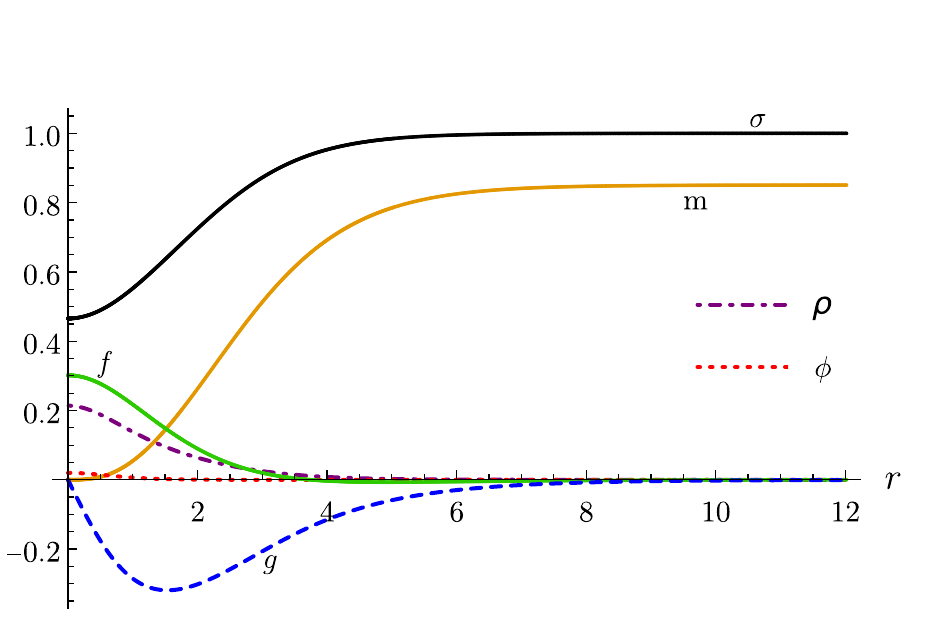}
		\qquad
	 	\includegraphics[scale=0.3]{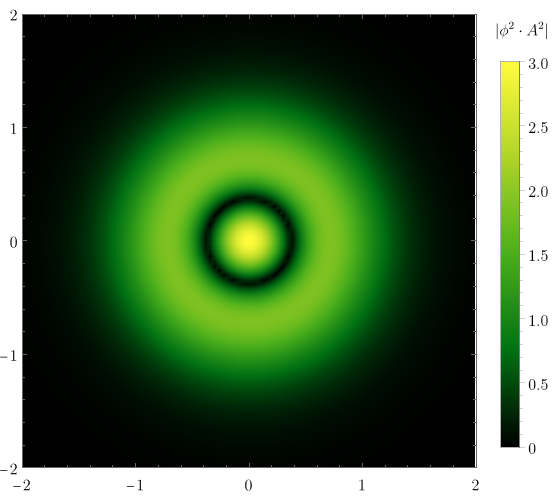}
	   	\caption{Non--minimally coupled SPS with $\alpha =43$, $\omega=10^{-6}$ and $\gamma/\sigma _0=1.750$ ($\mathcal{V}=0.698$), $\phi_0=0.020$, $M=0.851$ and $Q=0.571$. (Left) metric and matter functions radial profile: (solid black) metric function $\sigma$; (Solid yellow) matter function $m$; (Solid green) vector field function $f$; (Dot-dashed purple) energy density, $\rho$; (Dotted red) scalar field amplitude, $\phi$; And (dashed blue) vector field function $g$. (Right) $|\phi^2\mathbf{A}^2|$ (min--max normalized), black (yellow) color represents the absence (maximum value) of $|\phi^2\mathbf{A}^2|$. Observe that the solution is regular everywhere.}
	 	\label{fig:3.10}
	\end{figure}

%
	So far the interaction coupling was kept fixed while changing $\gamma/\sigma_0$. It is also interesting to consider SBSs with fixed $\phi_0$ (say) and examine the effects of varying $\alpha$. \autoref{fig:3.9} shows how $\omega$ depends on $\alpha$ for different values of $\phi_0$. Starting from $\alpha=-10$ and for small values of $\phi_0$, the frequency increases as $\alpha$ increases, reaching a maximum and then decreasing towards $\omega=0$ at $\alpha\approx43$. This behavior suggests the existence of configurations featuring static ($\omega=0$) scalar fields in equilibrium with oscillating ($\gamma\neq0$) vector fields. In fact, the condition $\omega=0$ does not spoil the boundary conditions at both the origin and infinity -- see~\autoref{sec:2.3}. Another evidence for their existence is the regularity of the matter functions of a SPS with a \textit{quasi--static} scalar field, as shown in~Fig.~\ref{fig:3.10}.

	We can observe that, as $\omega \to 0$, the SPS seems to be regular everywhere and no strange behavior of the matter nor metric functions exist (in opposition to the spontaneous matterized solutions). While the limit $\omega \to 0 $ (\textit{aka} real scalar field solutions) seem to be important, no distinct behavior exists.

%
\section{Conclusion and future work}\label{sec:4}

	This paper reported the existence of macroscopic self--gravitating Bose--Einstein condensates consisting of two equal--mass bosonic fields (a scalar and a vector) minimally coupled to Einstein's gravity and either minimally or non--minimally coupled to each other -- here dubbed \textit{scalaroca stars}. They can be thought of as (non--linear) superpositions of single--field BSs. 
	
	The family of minimally coupled ($\alpha=0$) solutions is likely one of the simplest examples of multi--field self--gravitating solitons in the literature. These stars were constructed for different frequencies $\omega$ and $\gamma$, and different masses $M$. For fixed values of $\gamma/\sigma_0$ and varying $\omega$, we have obtained families of solutions with different masses that branch out from the PS family and connect to the SBS family. These solutions tend to have lower mass than the corresponding PS but higher mass than the corresponding SBS.
	
	For the non--minimally coupled cases ($\alpha\neq0$), we first considered a positive coupling for both synchronized and non--sychronized frequencies. For the non--synchronized ($\omega\neq\gamma$) configurations, we fixed $\gamma/\sigma_0$ and varied $\omega$, just like in the minimally coupled case. The obtained solutions tend to exhibit similar qualitative behavior to the minimally coupled case for low values of $\alpha$. But, for larger values of $\alpha$, new branches of solutions appear, some of which can have higher mass than the PSs they branch out from. For the synchronized ($\omega=\gamma$) configurations, both frequencies are fixed and instead we change $\alpha$. Larger values of $\alpha$ tend to correspond to families with higher frequency. Also, unlike the non--synchronized case, the family of solutions that branch out from the PS solution does not join the SBS line.

	Lastly, we considered the effects of a negative coupling $\alpha<0$. The domain of existence is very similar to the positive coupling case, but the radial profile of the metric and matter functions can differ significantly.	
	
	Extensions of the work presented herein are manifold. A straighforward follow--up to it is to consider non--equal--mass bosonic fields (partially addresed in~\cite{Ma:2023vfa}). The existence of SPSs is expected to be very sensitive to the ratio between their bare masses, especially when the fields only interact gravitationally. It would be interesting to see if they exist solely for a finite range of this ratio. If so, the size of such interval may vary significantly with the coupling strength $\alpha$.
	
	Another possible add--on to this work is to construct and study \textit{rotating scalaroca stars}. Although they are likely to exist, just like the rotating counterparts of single--field BSs, it remains unclear how the addition of rotation affect SPSs. Furthermore, since spherically--symmetric configurations admit static scalar fields, it would be interesting to see whether rotating PSs in equilibrium with non--rotating BSs exist or not.
	
	Yet another possible line of research is to consider  adding an event horizon to these rotating configurations. While spherically symmetric black--hole solutions do not exist in this model\footnote{Assuming the existence of an event horizon at $r=r_H>0$, \textit{i.e.} $N(r_H)=0$, it can be shown that a finite energy density requires \textit{both} $f$ and $\phi$ to vanish at $r=r_H$. However, if $\phi(r_H)=0$, $\phi^{(n)}(r_H)=0$, $\forall n\in\mathbb{N}$, which means that $\phi$ is not analytic at the event horizon.}, rotating SPSs are expected to allow for black--hole generalizations. Indeed, black holes gravitationally bound to rotating single--field BSs are known to exist~\cite{Herdeiro:2014goa,Herdeiro:2016tmi}.
	
	Finally, of paramount importance is an in-depth analysis of the (linear and non--linear) stability of SPSs. It was shown that some familes of solutions connect stable PSs to unstable SBSs, which raises the question whether their non--linear combination yields a stable or an unstable configuration. A look into their time evolution or their spectrum of quasi-normal modes could clarify this point and reveal if new modes (\textit{i.e.} absent in the spectra of single--field BSs) appear.

\section*{Acknowledgements}
%
The authors would like to thank Carlos Herdeiro, Eugen Radu and Ippocratios Saltas for their valuable comments on an earlier version of this manuscript. 

Alexandre M. Pombo is supported by the Czech Grant Agency (GA\^CR) under the grant number 21-16583M. N. M. Santos is supported by the FCT grant SFRH/BD/143407/2019. This work is supported by the Center for Research and Development in Mathematics and Applications (CIDMA) and the Centre of Mathematics (CMAT-UM) through the Portuguese Foundation for Science and Technology (FCT -- Funda\c{c}\~ao para a Ci\^encia e a Tecnologia), references UIDB/04106/2020, UIDP/04106/2020, UIDB/00013/2020 and
UIDB/00013/2020, and by national funds (OE), through FCT, I.P., in the scope of the framework contract foreseen in the numbers 4, 5 and 6 of the article 23, of the Decree-Law 57/2016, of August 29, changed by Law 57/2017, of July 19. The authors acknowledge support from the projects CERN/FIS-PAR/0027/2019, PTDC/FIS-AST/3041/2020 and CERN/FIS-PAR/0024/2021. This work has further been supported by  the  European  Union's  Horizon  2020  research  and  innovation  (RISE) programme H2020-MSCA-RISE-2017 Grant No.~FunFiCO-777740 and by the European Horizon Europe staff exchange (SE) programme HORIZON-MSCA-2021-SE-01 Grant No. NewFunFiCO-101086251.

\bibliographystyle{jhep}  
\bibliography{references}

\end{document}